\documentclass[aps,pra,twocolumn,floatfix,showpacs]{revtex4-1}
\usepackage{graphicx}
\usepackage{amssymb}
\usepackage{amsmath}
\usepackage{setspace}
\usepackage{comment}
\usepackage[normalem]{ulem}

\usepackage[dvipsnames]{xcolor}
\usepackage{soul}
\usepackage[bottom]{footmisc}
\usepackage{color}

\newcommand{\fix}[1]{\textcolor{black}{#1}}
\newcommand{\lfix}[1]{\textcolor{black}{#1}}
\newcommand{\gfix}[1]{\textcolor{black}{#1}}
\newcommand{\hfix}[1]{\textcolor{black}{#1}}

\newcommand{\rfix}[1]{\textcolor{black}{#1}}
\newcommand{\rcfix}[1]{\textcolor{black}{#1}}

\begin{document}

\title{Photoionization of Rydberg Atoms in Optical Lattices}
\author{R.~Cardman}
\email{rcardman@umich.edu}
\author{J.~L.~MacLennan}
\author{S.~E.~Anderson}
\author{Y.-J.~Chen}
\altaffiliation[Present address:]{ National Institute of Standards and Technology, Boulder, Colorado 80305, USA}
\author{G.~Raithel}
\affiliation{Department of Physics, University of Michigan, Ann Arbor, MI 48109}
\date{\today }

\begin{abstract}
We develop a formalism for photoionization (PI) and potential energy curves (PECs) of Rydberg atoms in \gfix{ponderomotive} optical lattices and apply it to examples covering several regimes of the optical-lattice depth. The effect of lattice-induced PI on Rydberg-atom lifetime ranges from noticeable to highly dominant when compared with natural decay. The PI behavior is governed by the generally rapid decrease of the PI cross sections as a function of angular-momentum ($\ell$), \hfix{ lattice-induced $\ell$-mixing across the optical-lattice PECs, and interference of PI transition amplitudes from the lattice-mixed into free-electron states.} In GHz-deep lattices, $\ell$-mixing leads to a rich PEC structure, and the \rfix{significant} low-$\ell$ PI cross sections are distributed over many lattice-mixed Rydberg states. In lattices less than several tens-of-MHz deep, atoms on low-$\ell$ PECs are essentially $\ell$-mixing-free and maintain large PI rates, while atoms on high-$\ell$ PECs trend towards being PI-free. Characterization of PI in GHz-deep Rydberg-atom lattices may be beneficial for optical control and quantum-state manipulation of Rydberg atoms, while data on PI in \rfix{shallower} lattices are potentially useful in high-precision spectroscopy and quantum-computing applications of lattice-confined Rydberg atoms.
\end{abstract}

\maketitle
\section{Introduction}
Rydberg atoms in optical lattices and traps have gained interest in the fields of quantum computing and simulations~\cite{Zhang.2011a,Nguyen.2018,
Barredo.2020, Wilson.2019}, quantum control~\cite{Cardman.2020a}, and high-precision spectroscopy~\cite{Moore.2015b, Ramos.2017, Malinovsky.2020a}, as the lattice confines the atoms and extends interaction times. However, the binding energy of Rydberg atoms is several orders of magnitude below the photon energy $\hbar \omega$ of commonly used optical-lattice fields. Optical photoionization (PI) of the Rydberg valence electron leads to lifetime reduction
and decoherence. \fix{Lattice-induced PI can broaden radio-frequency (RF) transitions between Rydberg states and limit the fidelity of Rydberg-atom quantum-control and -simulation schemes that involve coherences in the RF domain. The PI can also degrade optical coherences between ground and Rydberg states \gfix{that can be} induced by $\lesssim 1$-kHz-linewidth lasers. \gfix{Such lasers} are becoming more widely used in metrology~\cite{Nez.1993a,Zhang.2017a,Campbell.2017a} and may become useful in research involving long-lived, lattice-trapped Rydberg atoms.}
%As lasers with linewidths much smaller than typical Rydberg-state decay rates ($\sim 10^4$~s$^{-1}$) are becoming more widely
%used~\cite{Nez.1993a,Zhang.2017a,Campbell.2017a}, PI-induced broadening of Rydberg levels may also become a limiting factor in optical preparation and control of Rydberg atoms in lattices. 
A characterization of PI in Rydberg-atom optical lattices will be helpful for ongoing and emerging Rydberg-atom applications.

A Rydberg atom in a laser field is subject to the ponderomotive, $e^2 {\bf{A}}^2 (\textbf{r}) / (2 m_{\rm{e}})$, and the $e {\bf{A}} (\textbf{r}) \cdot {\bf {p}} / m_{\rm{e}}$ interactions, with $e$, $m_{\rm{e}}$, $\textbf{r}$, $\textbf{p}$, and ${\bf{A}} (\textbf{r})$ denoting the magnitude of the fundamental charge, electron mass, electron position and momentum in the laboratory frame, and the position-dependent vector potential of the field, respectively~\cite{Friedrichbook}. \rfix{Interplay between these two interactions has previously been discussed in Refs.~\cite{Bucksbaum.1986,Bucksbaum.1987,Eberly.1986} in the context of above-threshold ionization}. In an inhomogeneous light field, such as an optical lattice, the ponderomotive ${\bf{A}}^2$ term generates an optical force on the Rydberg electron that depends on the intensity gradient \gfix{of the optical-lattice interference pattern and its overlap with} the spatial distribution of the Rydberg-electron wavefunction. \rfix{Effects of the ponderomotive force on free electrons in a standing-wave laser field were studied before in Refs.~\cite{Bucksbaum.1988,Freimund.2001}. \gfix{The Rydberg} electron is quasi-free, allowing the ponderomotive force to enable optical-lattice traps for Rydberg atoms~\cite{SEAnderson.2011, Lampen.2018}.}
%Cite Kuzmich? Sarah's? with more Google citations, take maybe 2 of ours and also some recent ones
The spatial period of Rydberg-atom optical lattices, which is on the order of the laser wavelength $\lambda$, is similar to the diameter of the trapped atoms, a situation that differs from most optical lattices, in which the atoms are point-like relative to $\lambda$. A ponderomotive optical lattice couples Rydberg states over a wide range of electronic angular momenta, $\ell$, ~\cite{Knuffman.2007, Younge.2009a}, affording capabilities in high-$\ell$ Rydberg-state initialization~\cite{Cardman.2020a, Younge.2009a} and Rydberg-atom spectroscopy free of selection rules for $\ell$~\cite{Knuffman.2007, Moore.2015a}.
\hfix{In the present analysis, we expand upon earlier work by including lattice-induced Rydberg-atom PI in ponderomotive optical lattices with strong $\ell$- and $j$-mixing. Our model describes PI-induced decay in the lattice, as required, for instance, the aforementioned quantum-control and computing applications.}

%\rcfix{However, none of these references provide the simultaneous treatment of $\ell$-state mixing within the hydrogenic manifold by the pondeormotive effect and decay to free-electron states by lattice-induced PI. This system involving both mechanisms has never been explored in an experimental or theoretical study and is the subject of this work.}
%1
%One-photon recoil hbar*hbar*k^2/(2m)= 2pi * 2.027 kHz for Rb-87
%3GHz ~ 1.48*10^6 recoils
%20MHz ~9867 recoils

%An important effect of the lattice field is photoionization.
%Unlike the ponderomotive force, which results from the ${\textbf{A}}^2$-term in the atom-field interaction,
Optical and black-body-radiation-induced PI result from the $\textbf{A}\cdot\textbf{p}$-term~\cite{Friedrichbook}.
Here we investigate laser-induced PI of Rydberg atoms trapped in an optical lattice.
%The paper is organized as follows.
In Sec.~\ref{PIOL}, we derive PI cross sections and rates for Rydberg atoms in plane-wave light fields and extend the results to Rydberg atoms in optical lattices. In Sec.~\ref{sec-PECs}, we obtain equations for the potential energy curves (PECs), the adiabatic Rydberg states, and their PI-induced decay rates in the lattice. In the examples in Sec.~\ref{app}, we focus on rubidium Rydberg atoms in a one-dimensional lattice formed by counter-propagating laser beams of 1064~nm wavelength. The lattice strength is characterized by the magnitude of the ponderomotive interaction relative to the unperturbed Rydberg-level separations. We present results for PECs and lattice-induced PI of $\ell$-mixed Rydberg atoms in a strong optical lattice, and of Rb~$50F$ atoms in a weaker, $\ell$-mixing-free optical lattice.
In the Appendix, we discuss fundamental aspects of optical PI of Rydberg atoms.

\section{PI of Rydberg atoms}
\label{PIOL}

%For the Rydberg-atom optical lattices that will be discussed in Sec.~\ref{app}, the effect of PI is most conveniently accounted for using PI cross sections. In this section, we provide  equations for PI matrix elements and how these translate into PI cross sections.

\subsection{Basic PI cross sections}

The lowest-order transition rate between atomic states due to
an interaction $\hat{H}_{\rm int}$ is given by Fermi's golden rule, $\Gamma=\frac{2\pi}{\hbar}|\langle f|\hat{H}_{\rm int}|i\rangle|^{2}\rho(\epsilon)$, with final-state energy $\epsilon$ and density of final states $\rho(\epsilon)$ ~\cite{Friedrichbook}. For PI in a plane-wave field, it is
$\hat{H}_{\rm int} = e \hat{\bf{A}}  \cdot \hat{\bf {p}} / m_{\rm{e}}$, and the final state $\vert f \rangle$ is
a free-electron state. We normalize the free-electron states per unit energy, {\sl{i. e.}}  $\langle f' \vert f \rangle = \delta(\epsilon' -\epsilon) \delta_{\eta',\eta}$, with $\eta$ denoting the angular-momentum quantum numbers ($\ell,m_{\ell}$) and $\rho(\epsilon)$ being equal to 1 per unit energy.
The PI cross section $\sigma_{\rm PI}$ is determined by dividing the PI rate by the photon flux density, $I/(\hbar\omega)$, where $I$ is the field intensity and $\omega$ its angular frequency. In SI units, for a linearly polarized field (polarization unit vector $\hat{\textbf{n}}$) with wave vector $\textbf{k}$, the PI cross section is (see Appendix~\ref{atomfield})
\begin{equation}\label{PIcross}
\sigma_{\rm PI}=\frac{\pi e^{2}\hbar^{2}}{\epsilon_{\rm 0} m_{\rm e}^{2} \omega c} \left|\hat{\textbf{n}}\cdot\int \psi_{f}^{\ast} \medspace e^{i\textbf{k}\cdot\textbf{r}_e} \medspace \nabla_{e} \psi_{i} \medspace d^{3}r_e \right|^{2}\left(\frac{1}{E_{H} \, a_0^2}\right) \, ,
\end{equation}
where $\textbf{r}_e$ denotes the relative Rydberg-electron coordinate, $E_H$ the atomic energy unit, \rcfix{$\psi_{i}$ and $\psi_{f}$ are the initial- and final-state wavefunctions, respectively,} and $a_0$ \rcfix{is} the Bohr radius. The last term converts the squared matrix element, which is in atomic units, into SI units. For PI of Rydberg atoms the electric-dipole approximation (EDA) typically is valid, as shown in~\cite{Anderson.2013a} \rcfix{in the context of a shallow optical lattice with no $\ell$-mixing. The validity of the EDA is} discussed in greater detail in Appendix~\ref{noapprox}. The EDA is implemented by setting $e^{i\textbf{k}\cdot\textbf{r}_e} = 1$ in Eq.~\ref{PIcross}. The resultant expression for the matrix element is referred to as ``velocity form'', used throughout this paper to compute the PI cross sections.

For $\hat{\bf{p}}$-independent atomic potentials, the matrix element in Eq.~\ref{PIcross}, with the EDA applied, can be transformed into ``length form'', leading to
\begin{equation}\label{PIcrossL}
\sigma_{\rm PI,L}=\frac{\pi e^{2} \omega}{\epsilon_{\rm 0} c  } \medspace \left|\hat{\textbf{n}}\cdot\int \psi_{f}^{\ast} \medspace {\bf{r}}_e \medspace \psi_{i} \medspace d^{3}r_e\right|^{2}\left(\frac{a_{\rm 0}^{\rm 2}}{E_{H}} \right) \, .
\end{equation}
This length-form expression for the PI cross section is not accurate if the atomic potential is $\ell$-dependent, as in the present work on Rb. 

%We therefore use Eq.~\ref{PIcross}, with $e^{i\textbf{k}\cdot\textbf{r}_e} = 1$. %Results are applicable to PI of Rydberg atoms with one valence electron and $\ell$-dependent Rydberg-electron potentials.

In the following, \hfix{we first consider PI of spin-less Rydberg basis states $\vert n, \ell, m_\ell \rangle$ into free states $| \epsilon',\ell', m'_\ell \rangle$. There,}  $n$ denotes the bound-state principal quantum number, $\epsilon'$ the free-electron energy, and $\ell_{>}$ the larger of the bound- and free-electron angular momenta, $\ell$ and $\ell'$. The shell-averaged PI cross section, \gfix{given by the average of the PI cross sections of the $m_\ell$-sublevels of the Rydberg state}, is
\begin{equation}\label{sigma_av}
\bar{\sigma}_{n, \ell}^{\epsilon', \ell'}=\frac{\pi e^{2} \hbar^2 }{3\epsilon_{\rm 0} m_{\rm{e}}^2 \omega c}\frac{\ell_{>}}{(2\ell+1)}|M|^{2}\left(\frac{1}{E_{H} \it a_{\rm 0}^{\rm 2}}\right),
\end{equation}
\noindent where $M$ is the radial part of the matrix element from Eq.~1 in atomic units, which is, with the EDA applied,
\begin{equation}\label{sigma_av2}
M = \int_{0}^\infty u_{\epsilon',\ell'}(r_e)\left[
u'_{n,\ell}(r_e) \mp \frac{u_{n,\ell}(r_e)}{r_e}\ell_{>} \right] dr_e.
\end{equation}
%I am getting M_{r} = \int_{0}^\infty u_{\epsilon',\ell'}(r_e)\left[ u'_{n,\ell}(r_e) \mp \frac{u_{n,\ell}(r_e)}{r_e}(\ell_{>}\mp 1) \right] dr_e.
There, \gfix{the upper sign is for $\ell_> = \ell'$, the lower sign for $\ell_> = \ell$, and $\ell'=\ell \pm 1$}. The functions $u_{*,\ell}(r_e)$ are given by $u_{*,\ell}(r_e) = r_e R_{*,\ell}(r_e)$, where $R_{*,\ell}(r_e)$ is the usual radial wavefunction, and $* = n$ or $\epsilon'$ for bound- and free-electron states, respectively. 

To illustrate the general behavior of PI cross sections of Rydberg states, we calculate
$\bar{\sigma}_{n, \ell}^{\epsilon', \ell'}$,
for a wide range of bound states $(n,\ell)$ and both PI channels $\ell' = \ell \pm 1$. The free-electron energy in atomic units is
\begin{equation}
\epsilon'=\frac{2 \pi \, a_0}{\alpha \lambda} - \frac{1}{2 n^{*2}} \quad,
\nonumber
\end{equation}
with the laser wavelength $\lambda$ in meters, the fine structure constant $\alpha$, and the effective quantum number of the Rydberg state, $n^*$. For the calculation of the bound-state and free-electron wavefunctions~\cite{Reinhard.2007a}, we use model potentials from~\cite{Marinescu.1994}, which have previously been employed to compute polarizabilities~\cite{Marinescu.1994c} and two-photon excitation rates~\cite{Marinescu.1994b} in Rb.
\fix{A table of the calculated $\bar{\sigma}_{n, \ell}^{\epsilon', \ell'}$, for $\lambda=1064$~nm, is provided as Supplementary Material}.

% Supplement file: Supp1_Data.dat

%The PI cross sections for $m_\ell$- and polarization-specific cases then follow from Eqs.~\ref{sigmazm} and~\ref{sigmaxm}, yielding cross sections $\sigma_{*,n,l,m_\ell}^{\epsilon',\ell'}$, where the subscript $* = z$ or $x$ denotes the light polarization, the other subscripts the bound-state quantum numbers, and superscripts the free-electron quantum numbers.

\begin{figure}
\begin{centering}
\includegraphics[width=3in]{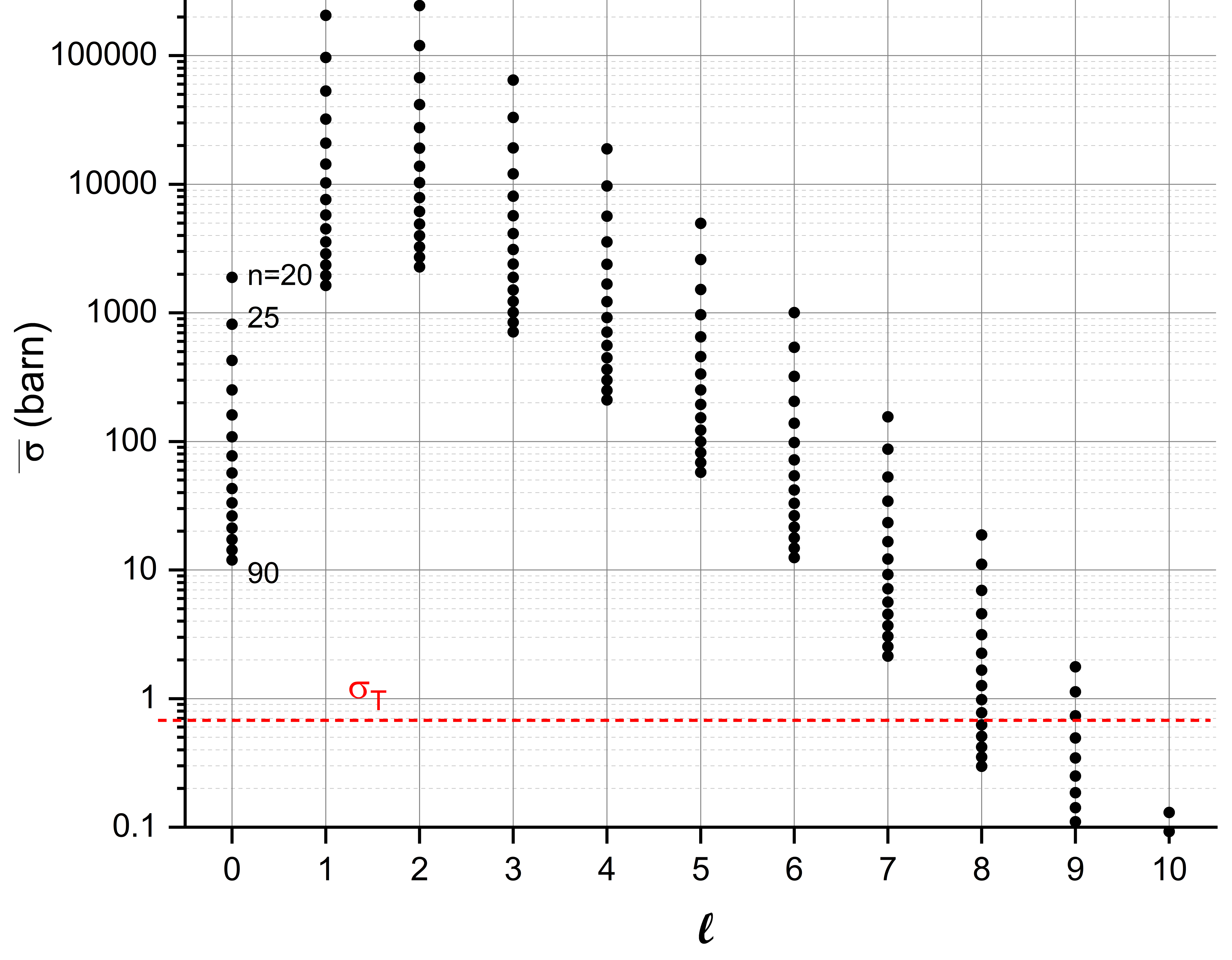}
\caption{Total shell-averaged PI cross sections $\bar{\sigma}_{n, \ell}$ for Rydberg $n$ and $\ell$ states of Rb for $\lambda =1064$~nm,
obtained by summing Eq.~\ref{sigma_av} over $\ell'$. For each $\ell$, the $n$ values range from 20 to 90 in steps of 5. The dashed line represents the Thomson cross section, $\sigma_{T}=$0.665 barn.}\label{sigma}
\end{centering}
\end{figure}

In Fig.~\ref{sigma} we show results for Rb in a $\lambda=1064$-nm field as a function of $n$ and $\ell$. The cross sections are generally quite large for low $\ell$, with an exception for the $S$-states that is caused by a Cooper minimum~\cite{Cooper.1962, Zatsarinny.2010}.
The calculated PI cross sections decrease rapidly as $\ell$ increases. For $\ell \gtrsim 10$, they drop below the elastic photon scattering cross section, given by the Thomson cross section, $\sigma_T=0.665$~barn. PI cross sections $\lesssim \sigma_T$ are likely too small to cause observable effects in applications.

\hfix{Since the lattice light has well-defined linear polarization, we note that for $z$-polarized} light the PI cross section for an atom in magnetic sublevel $m_\ell$ is
\begin{equation}\label{sigmazm}
\sigma_{z,n,\ell,m_\ell}^{\epsilon',\ell'}
=\frac{3(\ell_{>}^{2}-m_\ell^{2})}{(2\ell_{>}+1)(2\ell_{>}-1)}\frac{(2\ell+1)}{\ell_{>}}
\bar{\sigma}_{n, \ell}^{\epsilon', \ell'} \quad ,
\end{equation}
with $\bar{\sigma}$ from Eq.~\ref{sigma_av}, \hfix{while for $x$-polarized light it is}
\begin{equation} \label{sigmaxm}
\sigma_{x,n,\ell,m_\ell}^{\epsilon',\ell'}
=\frac{3}{2}\frac{(\ell'(\ell'+1)+m_\ell^{2})}{(2\ell_{>}+1)(2\ell_{>}-1)}\frac{(2\ell+1)}{\ell_{>}}
\bar{\sigma}_{n, \ell}^{\epsilon', \ell'} \quad .
\end{equation}

%The $m'_\ell$-states, or $m'_\ell$-state superpositions, of the free-electron states in Eqs.~\ref{sigmazm} and~\ref{sigmaxm} follow from the respective laser polarizations.
%Averaging the cross sections over the initial-state $m_\ell$ value in a polarized field yields $\bar{\sigma}$,  as does summing over all allowed transitions for an initial state with fixed $m_\ell$ immersed into an unpolarized field.

\subsection{PI cross sections and rates of lattice-mixed states}
\label{sec-matel-OL}

\hfix{Due to lattice-induced Rydberg state mixing, lattice-trapped Rydberg atoms are coherent superpositions of numerous basis states. Also,} the fine structure must be included, because it can be on the order of or larger than the optical-lattice trap depth. 
\hfix{Eq.~\ref{PIcross} then has to be evaluated for the lattice-mixed states $\vert i \rangle = \sum_{n, \ell, j, m_j} c_{n, \ell, j, m_j} \vert n, \ell, j, m_j \rangle$, with the total-angular-momentum quantum numbers $j$ and $m_j=m_\ell + m_s$, and electron-spin magnetic quantum number $m_s$.}

\hfix{Here we adopt a geometry in which a pair of counter-aligned lattice beams propagate along $z$, and the linear lattice polarization is along $x$.
The PI then has matrix elements in the $(m_\ell, m_s)$-basis given by
\begin{eqnarray}
M_{n, \ell, m_\ell, m_s}^{\epsilon',\ell',m'_\ell,m'_s} & = &
\langle \epsilon',\ell',m'_\ell,m'_s \vert %\hat{\frac{\partial}{\partial x_e}} 
{\rm {i}} \hat{p}_{x,e}
\vert n, \ell, m_\ell, m_s \rangle \nonumber \\
~ & ~ & \times \delta_{m'_\ell, m_\ell \pm 1} \delta_{m'_s, m_s} \delta_{\ell', \ell \pm 1} \quad,
\end{eqnarray}
in atomic units and with the $x$-component of the electron momentum, $\hat{p}_{x,e}$. We have added Kronecker $\delta$'s to exhibit the PI selection rules. The matrix elements have a radial part given by Eq.~\ref{sigma_av2} and angular parts that follow from \cite{Bethebook} p. 254. The PI cross section for the lattice-mixed states then is 
\begin{eqnarray}\label{PIcross2}
\sigma_{\rm PI}  & = & \frac{\pi e^{2}\hbar^{2}}{\epsilon_{\rm 0} m_{\rm e}^{2} \omega c} \sum_{\ell', m'_\ell, m'_s}
\Big \vert \sum_{n, \ell, j, m_j, m_\ell, m_s}
M_{n, \ell, m_\ell, m_s}^{\epsilon',\ell',m'_\ell,m'_s}
\nonumber \\ 
~ & ~ & \times \,
c_{n, \ell, j, m_j} \, \langle j, m_j \vert m_\ell m_s \rangle \Big \vert^{2} \left(\frac{1}{E_{H} \, a_0^2}\right) \, .
\end{eqnarray}
Note $M$ is in atomic units, and the term in () converts the matrix-element-square into SI units. Due to symmetry $m_j$ is fixed.
Since the lattice induces $\ell$- and $j$-mixing, 
the PI cross sections exhibit quantum interference in the inner sum, caused by the fact that several PI channels can lead from multiple basis states $\vert n, \ell, j, m_j \rangle$ into the same free state $\vert \epsilon', \ell', m'_\ell, m'_s \rangle$. }

\hfix{For given $\sigma_{\rm PI}$, the atom PI rate follows from
\begin{equation}
\label{eq:rate}
\Gamma =  I  \frac{\sigma_{\rm PI}}{\hbar \omega} \quad. 
\end{equation}
%Optical lattices are formed by the interference of multiple laser beams, leading to spatially varying intensity distributions and sometimes polarization gradients. The resultant light-shift forces are widely used for atom trapping, including Rydberg-atom trapping \lfix{\cite{Dutta.2000a, SEAnderson.2011, Zhang.2011a}}.
Since in the optical lattice the intensity $I$ varies within the volume of the Rydberg atom, it is not immediately obvious what to use for $I$ in Eq.~\ref{eq:rate}.} In fact, the atomic volume can extend over several nodes and anti-nodes of the light field~\cite{Dutta.2000a, SEAnderson.2011, Zhang.2011a}. The lattice-intensity variation within the atomic volume is important for the potential energy curves (PECs) and state-mixing in the lattice, as discussed in the next section. Our analysis given in the Appendix shows that the PI rates of Rydberg states are determined by the intensity at the exact CM location of the Rydberg atom, $I({\bf{R}}_0)$. We enter $I({\bf{R}}_0)$ into Eq.~\ref{eq:rate} to obtain the PI rates of the lattice-mixed Rydberg atoms. It is irrelevant how the field varies over the atomic volume. Especially \lfix{noteworthy is the fact that} the light intensity within the main lobes of the Rydberg electron wavefunction is not important. This finding is a consequence of the validity of the EDA for PI of Rydberg atoms, which is discussed in the Appendix. Laser-induced Rydberg-atom PI was \rfix{previously} measured in plane waves~\cite{Tallant.2010} and, in a spatially-sensitive manner, in an optical lattice~\cite{Anderson.2013a}.

\section{Potential energy curves}
\label{sec-PECs}

\subsection{Strong optical-lattice regime}
\label{subsec:SL}

Rydberg atoms in an optical lattice are subject to
both the ${\bf{A}} \cdot {\bf {p}}$ and the ponderomotive (${\bf{A}}^2$) interactions, giving rise to lattice-induced PI and PECs at the same time. In the following we describe our 
comprehensive formalism for both PI and PECs. In a one-dimensional optical lattice along the $z$-direction,
the PECs are calculated by first finding the Hamiltonian
 \begin{equation}\label{H}
 \hat{H}_{\rm lat}=\hat{H}_{\rm 0}+V_{\rm P}(\hat{z}_e+Z_{\rm 0})
 \end{equation}
\noindent
on a grid of fixed CM positions $Z_{\rm 0}$ of the atoms in the lattice. There, $\hat{H}_{\rm 0}$ is the field-free atomic Hamiltonian, and the operator $\hat{z}_e$ represents the relative $z$-coordinate of the Rydberg electron. Further, $V_{\rm P}(z)=e^{2}E^{2}(z)/(4m_{\rm e}\omega^{2})$ is the free-electron ponderomotive potential that follows from the ${\bf{A}}^2$-interaction, $E(z)$ the total lattice electric-field amplitude, and $z=z_e+Z_{\rm 0}$ the $z$-coordinate of the Rydberg electron in the laboratory frame. Classically, the ${\bf{A}}^2$-term may be thought of as the time-averaged kinetic energy of the electron quiver in the lattice electric field at the optical frequency~\cite{Dutta.2000a}. In a one-dimensional lattice along $z$,
\begin{equation}\label{2V0}
V_{\rm P}(z)= V_0 (1+\cos(2kz))
\quad ,
\end{equation}
with the full free-electron potential depth $2 V_0$ and $k=2 \pi / \lambda = \omega/c$. For a pair of lattice beams with equal single-beam electric-field amplitude $E_0$ and equal linear polarization, it is $2 V_0 = e^2 E_0^2 / (m_e \omega^2)$. The potential $V_{\rm P}(z)$ introduces couplings that are free of selection rules for $\ell$~\cite{Younge.2009a,Wang.2016}.
From a perturbation-theory viewpoint, the Rydberg-atom lattice is strong if the lattice depth approaches the characteristic energy scale of the unperturbed Rydberg atom, i.e., if $2 V_0 \gtrsim s E_H/n^3$,  with 
%atomic energy unit $E_H$, principal quantum number $n$, and a 
scaling parameter $s \sim 0.1$ that depends on the quantum defects of the atom. In a strong lattice, the lattice-induced couplings approach or exceed the quantum-defect-induced energy gaps between low- and high-$\ell$ states, \rfix{ causing mixing among such states.} 

\hfix{The PECs, $W_k(Z_0)$, and the lattice-mixed  adiabatic Rydberg states $\vert \psi_k(Z_0) \rangle$
are found by solving 
\begin{equation}
\hat{H}_{\rm {lat}}(Z_{\rm 0}) \vert \psi_k(Z_0) \rangle =  W_k(Z_0)  \vert \psi_k(Z_0) \rangle ,
\end{equation}
with an index $k$ labeling the PECs and their adiabatc states. 
We use representations of the $\vert \psi_k(Z_0) \rangle$ in the basis of the field-free states $\vert n, \ell, j, m_j \rangle$ in Eq.~\ref{PIcross2} in order to yield the PI cross sections, $\sigma_k (Z_0)$, and the PI rates, $\Gamma_k(Z_0)$, from Eq.~\ref{eq:rate}. It is observed that
the PI rates $\Gamma_k$, which trace back to the
${\bf{A}} \cdot {\bf {p}}$ interaction, and the free-electron ponderomotive lattice shift, which arises from the ${\bf{A}}^2$ interaction, both scale with the
intensity at the atom's CM location,
\begin{eqnarray}
 \Gamma_{k} (Z_{\rm 0}) & = & I(Z_{\rm 0}) \frac{\sigma_k (Z_0)}{\hbar\omega} \nonumber
 \\                   V_{\rm P}(Z_0) & = & I(Z_{\rm 0}) \frac{e^2}{2 c \epsilon_{\rm 0} m_{\rm e} \omega^2} \quad.
 \label{eq:gamma}
\end{eqnarray}
}

We note that the PECs $W_k(Z_0)$ satisfy
\begin{equation}\label{Vad}
W_k(Z_0)=\int V_\mathrm{P}\left(z_e+Z_{\rm 0}\right)|\psi_k\left({\bf{r}}_e; Z_0 \right)|^2\medspace d^3 r_e,
\end{equation}
\noindent which represents a spatial average of $V_{\rm P}$, weighted by the wavefunction densities $|\psi_k\left({\bf{r}}_e; Z_0 \right)|^2$ of the adiabatic states $\vert \psi_k(Z_0) \rangle$. The wavefunction density is traced over the electron spin.
Since the $\vert \psi_k(Z_0) \rangle$ are not known before diagonalization of the Hamiltonian in Eq.~\ref{H}, Eq.~\ref{Vad} generally cannot be used to calculate PECs (exceptions are discussed in the next Sec.~\ref{subsec:WL}). Instead, the Hamiltonian in Eq.~\ref{H} must be diagonalized to simultaneously yield both the PECs, $W_k(Z_{\rm 0})$, and the  $\vert \psi_k (Z_0) \rangle$. \hfix{The latter then allows computation of $\sigma_k(Z_0)$.}

\subsection{Weak optical-lattice regime}
\label{subsec:WL}

If the Rydberg-atom lattice is weak, $2 V_0 < s E_H/n^3$, there are cases in which the ponderomotive potential $V_{\rm P}(z)$ does not cause lattice-induced state mixing of the unperturbed Rydberg levels. These cases include $nS_{1/2}$ Rydberg levels, and $nP_{j}$ and $nD_{j}$ levels if $2 V_0$ is also less than the fine structure splitting. For Rydberg states that are known to be mixing-free, the PECs can be obtained from
first-order non-degenerate perturbation theory,
\begin{equation}\label{Vad0}
W_k(Z_0)=\int V_\mathrm{P}\left(z_e+Z_{\rm 0}\right)|\psi_{k,0}\left({\bf{r}}_e\right)|^2\medspace d^3 r_e \quad .
\end{equation}
\noindent This expression amounts to a spatial average of $V_{\rm P}$, weighted by the wavefunction density of the unperturbed,  $Z_0$-independent state $\vert \psi_{k,0} \rangle = \vert n, \ell, j, m_j \rangle$,
\begin{eqnarray} \vert \psi_{k,0}({\bf{r}}_e) \vert^2 = |R_{n,\ell,j}(r_e)|^2 &
\, \Big[ &
\vert c_{\uparrow} \, Y_{\ell}^{m_j - 1/2} (\theta_e, \phi_e) \vert^2
\nonumber \\
~ & + &  \vert c_{\downarrow} \, Y_{\ell}^{m_j + 1/2} (\theta_e, \phi_e) \vert^2 \Big]
\quad , \nonumber
\end{eqnarray}
\hfix{with Clebsch-Gordon coefficients  $c_{\uparrow}$ and
$c_{\downarrow}$ for $m_s=1/2$ and $-1/2$, respectively,} and spherical Rydberg-electron
coordinates $(r_e, \theta_e, \phi_e)$. \gfix{The PEC index $k$ now
merely is} a shorthand label for the mixing-free state $\vert n,\ell,j,m_j \rangle$. PECs in weak lattices  have been investigated in Refs.~\cite{Younge.2010a,Anderson.2012a}. \hfix{Also, the PI cross section of $\vert n, \ell, j, m_j \rangle$ according to 
Eq.~\ref{PIcross2} greatly simplifies and there is no quantum interference of PI channels (as the inner sum has only one term for each $\vert \epsilon', \ell', m'_\ell, m'_s \rangle$). }

In certain scenarios, one can force applicability of non-degenerate perturbation theory by lifting degeneracies via application of an auxiliary DC electric or magnetic field, or a microwave field. If the auxiliary field suppresses lattice-induced state mixing, the adiabatic Rydberg states in the lattice become independent of $Z_0$, allowing a perturbative calculation of the PECs as in Eq.~\ref{Vad0}~\cite{Dutta.2000a, Ramos.2017}. In some of the cases, the fine-structure coupling can be lifted by the DC field, and the time- and $Z_0$-independent states become $\vert n, \ell, m_\ell, m_s \rangle$. In those cases, the wavefunctions to be used in Eq.~\ref{Vad0} are $\psi_{k,0} ({\bf{r}}_e) = \langle {\bf {r}}_e \vert n, \ell, m_\ell \rangle$, \hfix{and their PI rates are directly given by Eqs.~\ref{sigma_av}, \ref{sigma_av2}, \ref{sigmaxm} and~\ref{eq:rate}, and by (incoherently) summing the rates over $\ell'$.} One such example is the weak one-dimensional lattice of Rb $50F$-states with an external DC electric field, discussed in Sec.~\ref{app1}.

\section{Results}
\label{app}

%In this section we discuss examples of Rydberg-atom optical lattices, which are relevant to ongoing work elsewhere.
%Rydberg-atom lattices and traps recently gained interest as exemplary systems in the fields of quantum computing and simulations~\cite{Zhang.2011a,Nguyen.2018,Barredo.2020}, quantum control~\cite{Cardman.2020a}, and high-precision spectroscopy~\cite{Moore.2015a, Ramos.2017, Malinovsky.2020a}

\subsection{An implementation of a strong optical lattice}
\label{app2}

%Following the distinction made in Sec.~\ref{sec-PECs}, for $n \sim 50$ the strong-lattice condition $2 V_0 \gtrsim s E_H/n^3$ amounts to $2 V_0 \gtrsim h \times 5$~GHz, equivalent to $\sim 10^6$ single-photon recoil energies of Rb in a $1064$-nm lattice, $E_{rec} = h \times 2.027$~kHz.

\begin{figure*}[t]
\includegraphics[scale=0.325]{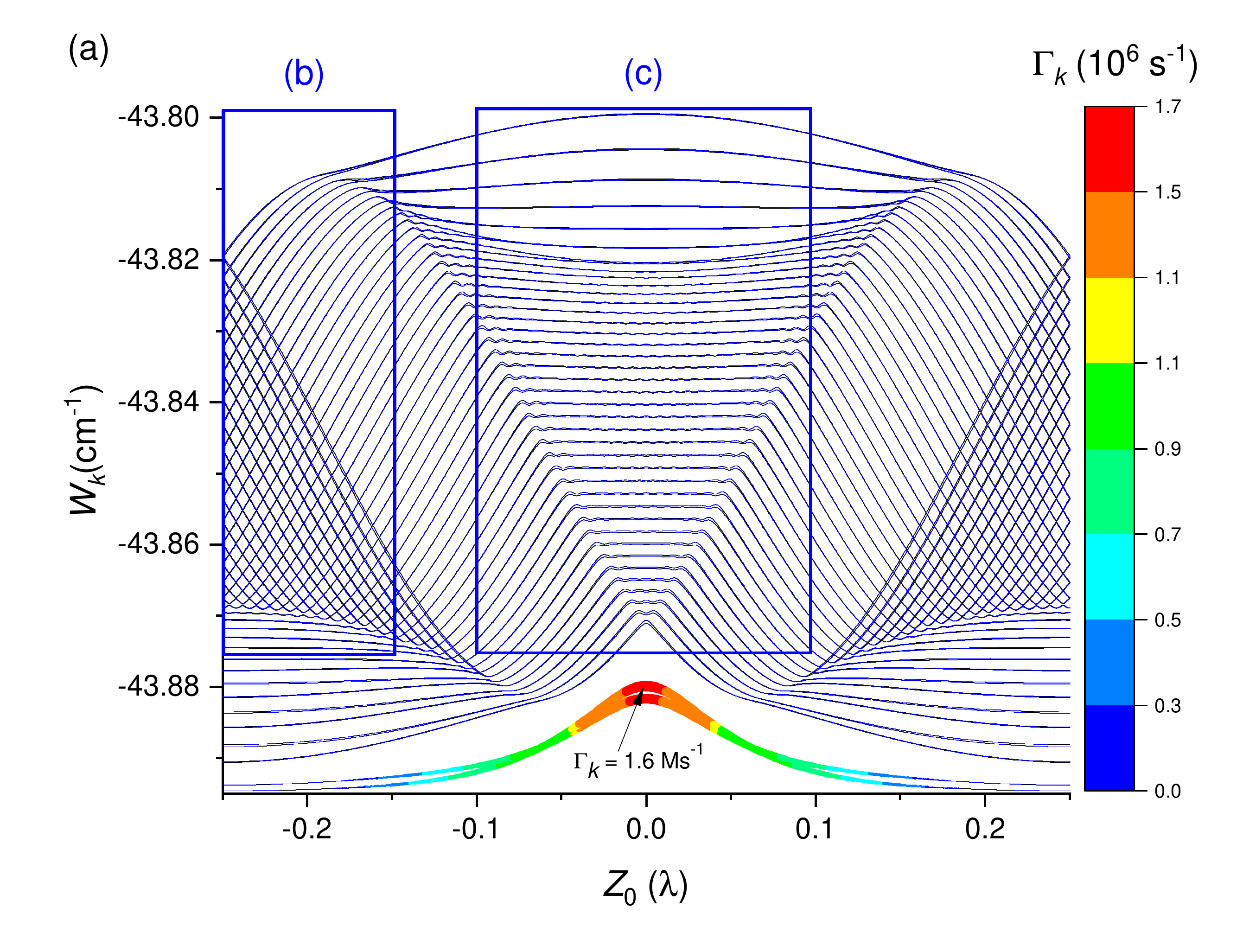}
\includegraphics[scale=0.325]{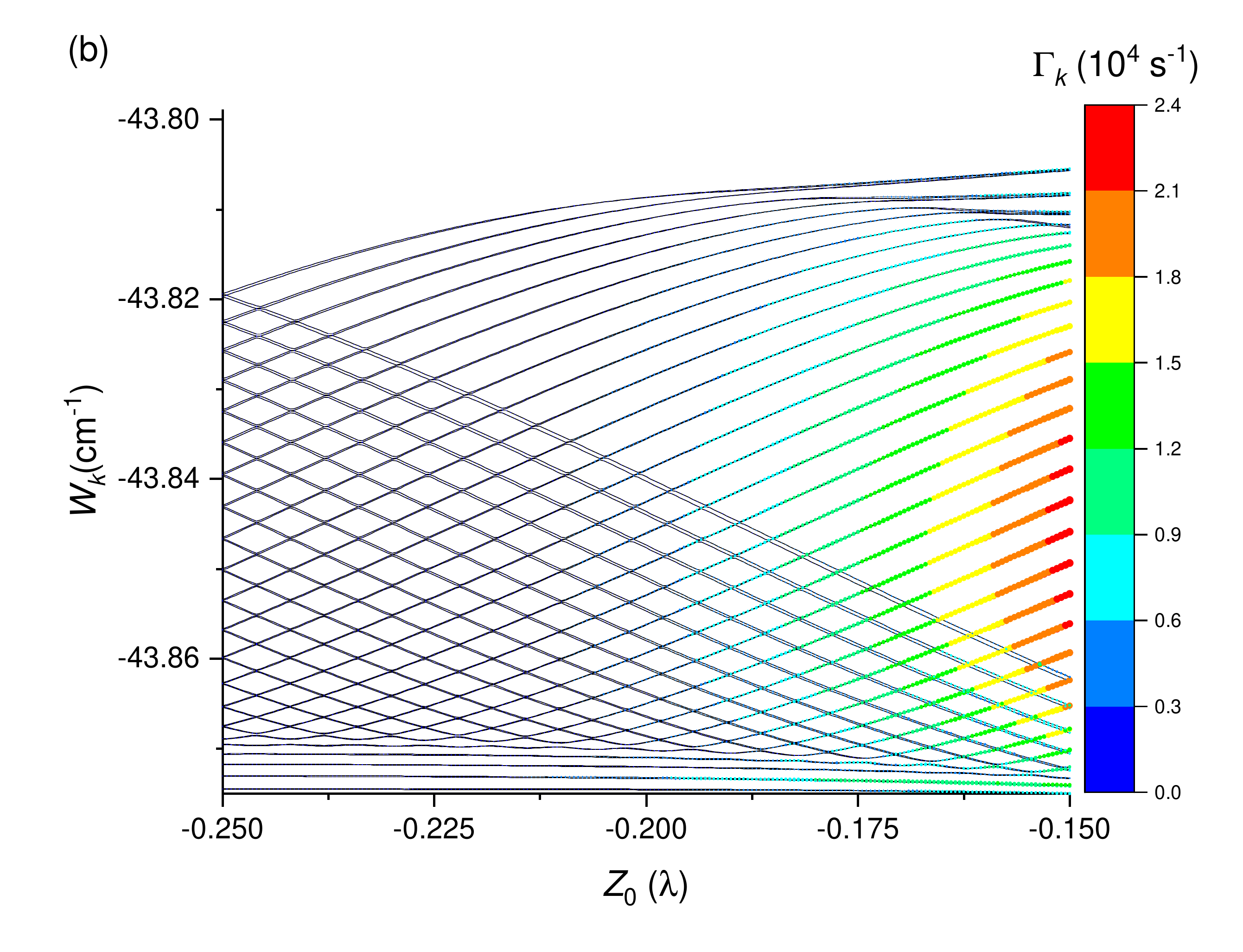}\\
\includegraphics[scale=0.325]{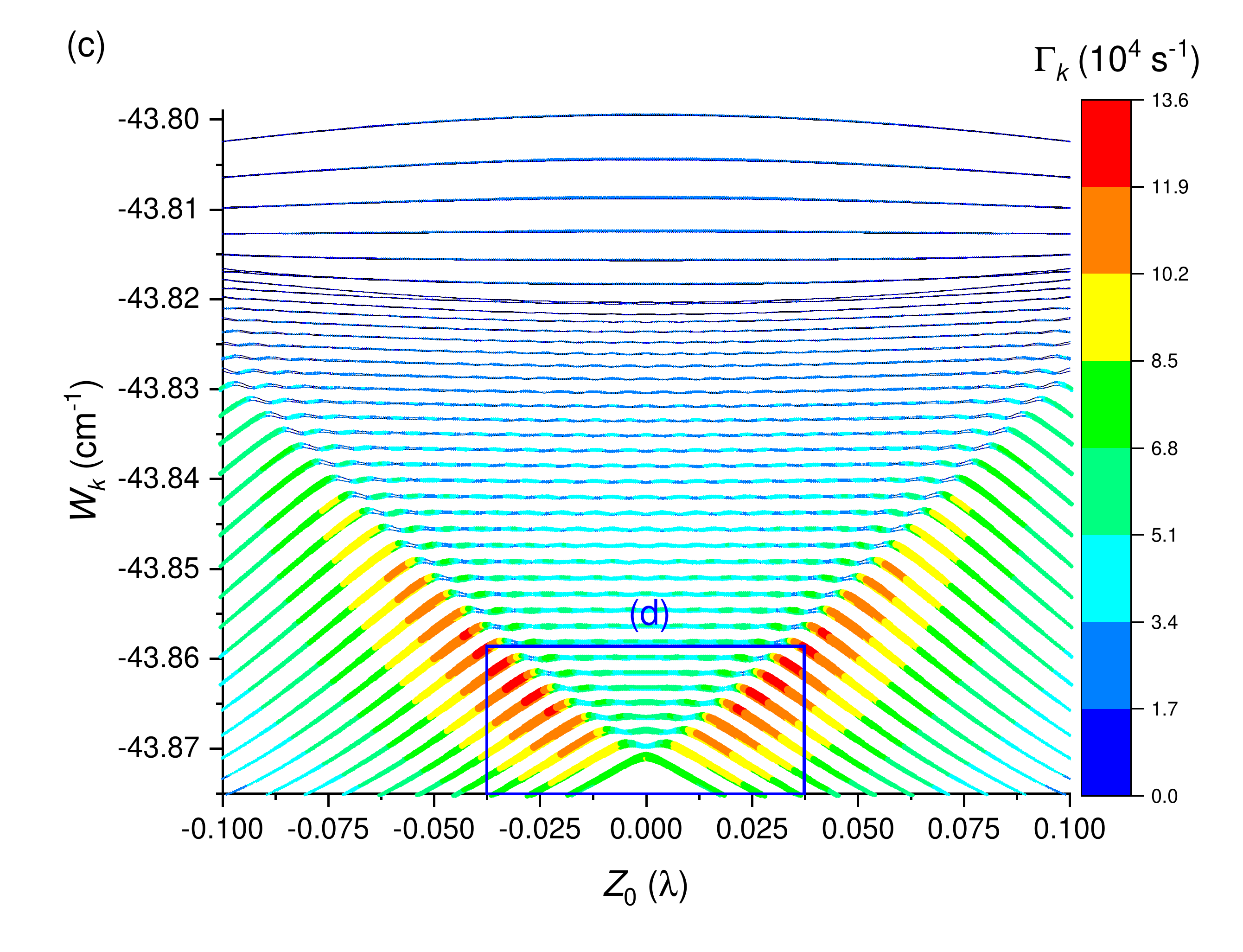}
\includegraphics[scale=0.325]{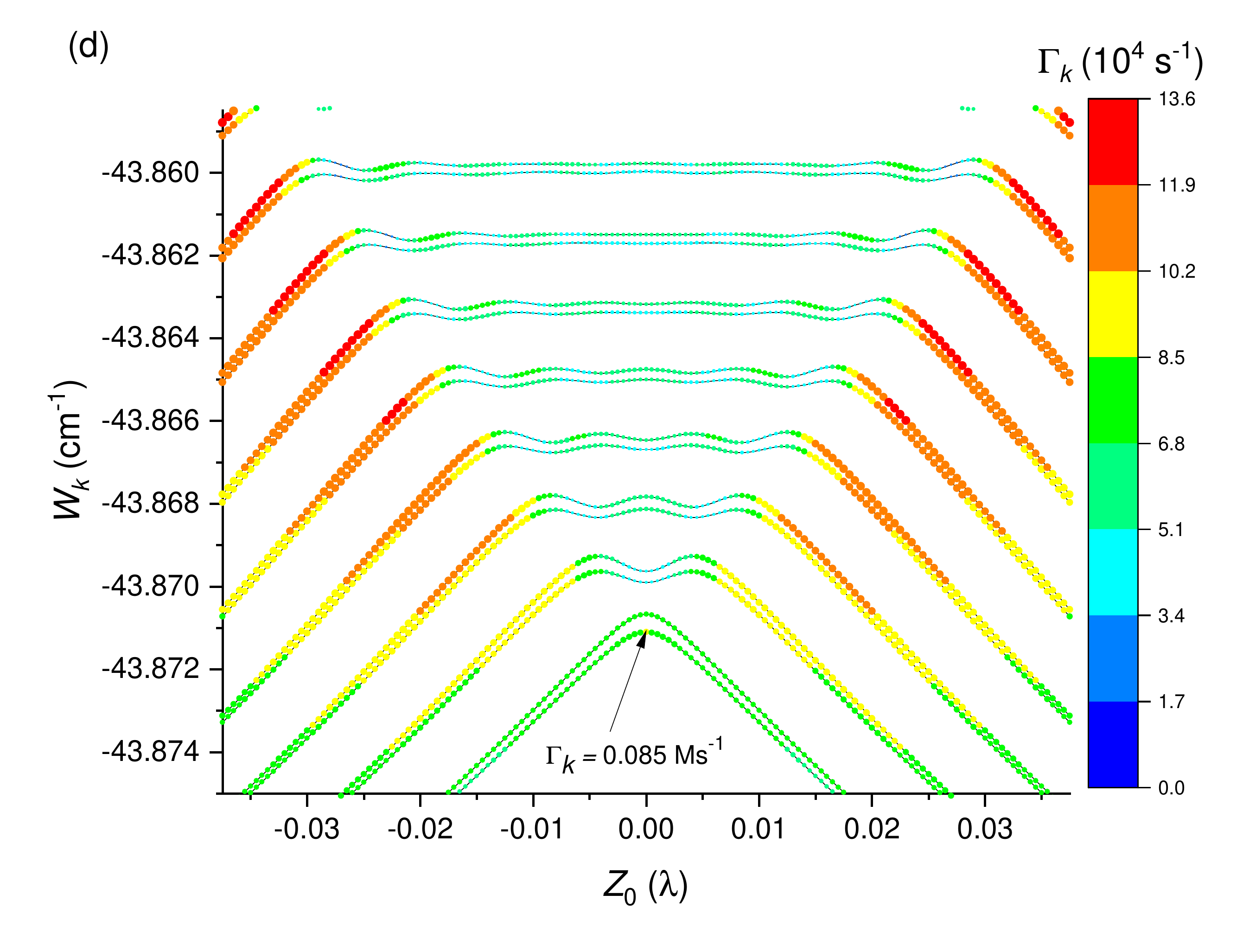}
\caption{PECs in a one-dimensional ponderomotive optical lattice of Rb Rydberg atoms for $n=50$ and $m_j = 1/2$, $\lambda=1064$~nm, and lattice depth $2V_0 = h \times 3$~GHz $=1.48\times10^{6} \, E_{rec}$. PEC energies are in cm$^{-1}$ relative to the ionization threshold, and CM positions $Z_0$ in units $\lambda$.
The boxed regions in (a) correspond to the full regions displayed in panels (b) and (c), while the boxed region in (c) corresponds to the full region displayed in panel (d). The color of the dots on the PECs shows PI rate $\Gamma_k(Z_0)$, on the color scales provided, and the dot diameter is proportional to $\Gamma_k(Z_0)$. For clarity, the dot diameters in (b) are enhanced by a factor of 50 relative to those in (a), and those in (c) and (d) by a factor of 10. The close-up view in (d) shows $\approx 10$-nm-period structures and a small fine-structure splitting of the PECs.}
\label{adpot}
\end{figure*}

In strong Rydberg-atom optical lattices, 
lattice-induced state mixing gives rise to a rich structure of PECs. 
This is illustrated in Fig.~\ref{adpot} for $n=50$, $m_{j}=1/2$, and lattice depth $2V_0 = h \times 3$~GHz, equivalent to $1.48\times10^{6} E_{rec}$, with the single-photon recoil energy of Rb for $\lambda = 1064$-nm, $E_{rec} = h \times 2.027$~kHz. For this lattice it is $2 V_0 \sim 0.1 E_H/n^3$, placing it in the strong-lattice regime as defined in Sec.~\ref{sec-PECs}. Fine structure and quantum defects~\cite{Gallagherbook} are included in the calculation. The diameters and colors of the dots on the PECs in Fig.~\ref{adpot} indicate the PI rates, $\Gamma_k(Z_0)$, of the PECs.

The lattice primarily mixes states of small quantum defects, which covers the vast majority of Rydberg states. The adiabatic states of the PECs,
$\vert \psi_k (Z_0) \rangle$, are coherent superpositions of a wide range of low-$\ell$ and high-$\ell$ states, including circular Rydberg states. The lowest-energy curves in Fig.~\ref{adpot}(a) are substantially perturbed $50F$-states, which are lowered in energy due to their quantum defect and are not entirely mixed into the manifold of high-$\ell$ states, which have near-zero quantum defect (states with $\ell \geq 4$ in Rb). The lattice-induced mixing of $F$-character into the high-$\ell$ states is efficient enough to
make the latter laser-excitable from a low-lying $D$-level.
For instance, the three-step excitation sequence $5S_{1/2}$ $\rightarrow$
$5P_{1/2}$ $\rightarrow$ $5D_{3/2}$ $\rightarrow$ $nF_{5/2}$ using 795~nm, 762~nm, and $\sim$1260~nm laser light would be suited for a spectroscopic study of these PECs.

Several prominent features of the PECs in Fig.~2 can be interpreted based on analogies with the Stark and diamagnetic effects in Rydberg atoms~\cite{Gallagherbook, Friedrichbook}. Near the inflection points of the lattice [$Z_{\rm 0}=\pm\lambda/8$ in Fig.~\ref{adpot}(a)], the PECs include sets of about 50 straight, parallel lines that resemble the level structure of the linear DC Stark effect. Since the ponderomotive potential $V_{\rm P}(z)$ is linear in these regions, the analogy with the DC Stark effect is  expected~\cite{Younge.2009a}.  Near the nodes and anti-nodes of the lattice [$Z_{\rm 0}=0,\pm\lambda/4$ in Fig.~\ref{adpot}(a)], the levels resemble the rotational and vibrational diamagnetic energy-level structure of Rydberg atoms~\cite{Zimmerman.1978a,Gay.1983a,Cacciani.1986a,vanderVeldt.1993a}. This similarity also is expected, because the ponderomotive potential near the nodes and anti-nodes and the diamagnetic potential share a quadratic dependence on position~\cite{Younge.2009a}. Spectroscopic studies in high-intensity lattices are yet to reveal the PEC structures shown in Fig.~\ref{adpot}.

The PI rates on the PECs, $\Gamma_k(Z_0)$, overall scale with the lattice intensity at the atomic CM location, \gfix{which is proportional to} $(1+\cos( 2 k Z_0 ))$. The maximum $\Gamma_k$-values in Fig.~\ref{adpot}(a) are $\Gamma_k \approx 1.6\times10^{6}$~s$^{-1}$ for the $50F$-like states at $Z_{\rm 0}=0$, where the lattice intensity is maximal. For the high-$\ell$ states within the range displayed in Fig.~\ref{adpot}(b), 
which is near a lattice-intensity minimum, the $\Gamma_k$ range between $2\times10^{4}$~s$^{-1}$ and zero (at the exact anti-node positions). For the high-$\ell$ states within the range of Fig.~\ref{adpot}(c), near an intensity maximum, the $\Gamma_k$-values peak at about $10^{5}$~s$^{-1}$. Since radiative decay rates and black-body-radiation-induced transition rates of Rydberg levels around $n=50$ are only on the order of $10^{4}$~s$^{-1}$, PI-induced decay in the lattice will be quite noticeable even for the high-$\ell$ states. For the $50F$-like states, it will greatly exceed natural decay, for conditions as in Fig.~\ref{adpot}.

\hfix{Due to the strong dependence of the PI cross sections on $\ell$, seen in Fig.~1, it is not obvious how much quantum interference of PI amplitudes from lattice-mixed states (inner sum in Eq.~\ref{PIcross2}) matters. The importance of interference can be assessed by taking an incoherent sum, in which the left vertical bar in Eq.~\ref{PIcross2} is moved to the inside of the inner sum, and by comparing coherent- with incoherent-sum results. The relative error in cross sections incurred by taking incoherent sums, averaged over all states $\vert \psi_k (Z_0) \rangle$ in Fig.~2, is 0.05, with a standard deviation of 0.04. While this error is too small to matter in cases where the lattice-induced PI rate simply has to be below an application-specific tolerance limit, it may be large enough to be noticeable in PI rate measurements.}

In possible future experimental work, an ultra-deep Rydberg-atom lattice with a
depth of $2V_0= h \times 3$~GHz, as considered in this section, could be achieved by focusing two counter-propagating laser beams, each with a power of 200~W, into a confocal 
spot with $w_{\rm 0}=20~\mu$m. Such a lattice can be prepared, for instance, by using a near-concentric field enhancement cavity~\cite{Chen.2014}, with the Rydberg atoms loaded into the focal spot of the cavity.

The PI-induced spectroscopic level widths in
Fig.~\ref{adpot}, which are $\Gamma_k/(2 \pi) \lesssim 250~$kHz, should be large enough to become visible in spectroscopic measurement of PECs with narrow-linewidth lasers (linewidth $\lesssim 100~$kHz).
%This measurement scenario} would involve laser excitation of laser-cooled atoms prepared in the lattice. In the case of multi-photon excitation through intermediate low-lying atomic states, such as the Rb~$5P_j$ and $5D_j$-levels, which have natural level widths of 6~MHz and 0.7~MHz, respectively, the  excitation would have to be sufficiently far-off-resonant from the intermediate states to avoid radiation pressure effects, level broadening caused by photon scattering from the intermediate states, and PI of intermediate states by the lattice laser.
Another possible measurement method for PEC curves and level widths would be microwave spectroscopy from a suitable low-$\ell$ launch Rydberg state. This method would essentially be Doppler-effect-free and benefit from the Hz-level linewidth of typical
microwave sources, resulting in higher spectral resolution.
However, it would add experimental complexity due to the need
to account for the PI and level shifts of the Rydberg launch state within the optical lattice.

%In another remark on possible experimental work, }
We note that near $Z_{\rm 0}=0$ and $\pm\lambda/4$ in Fig.~\ref{adpot}, and within certain energy regions, the PECs feature series of periodic wells with a periodicity of $\approx 10$~nm and a depth in the range of $h \times 10$ to 100~MHz. The periodicity is about a factor of 50 smaller than the fundamental $\lambda/2$-periodicity of the
optical lattice, while the depth allows about one to three quantum states of the CM motion in each well, with tunneling-induced well-to-well coupling. On a single PEC there are as many as about 20 small periodic wells, making the system conducive to studies of tunneling-induced quantum transport. Further, since CM momentum exchange between CM wavefunctions and periodic gratings scales with the inverse of the spatial period, the 10-nm-period PECs in Fig.~\ref{adpot} may also serve well as large-angle Bragg reflectors and beam splitters for Rydberg-atom CM wavefunctions, which could potentially become useful in atom-interferometry applications~\cite{Cronin.2009}.

\subsection{An implementation of a weak optical lattice}
\label{app1}

\begin{figure*}[t]
\begin{centering}
\includegraphics[width=5.5in]{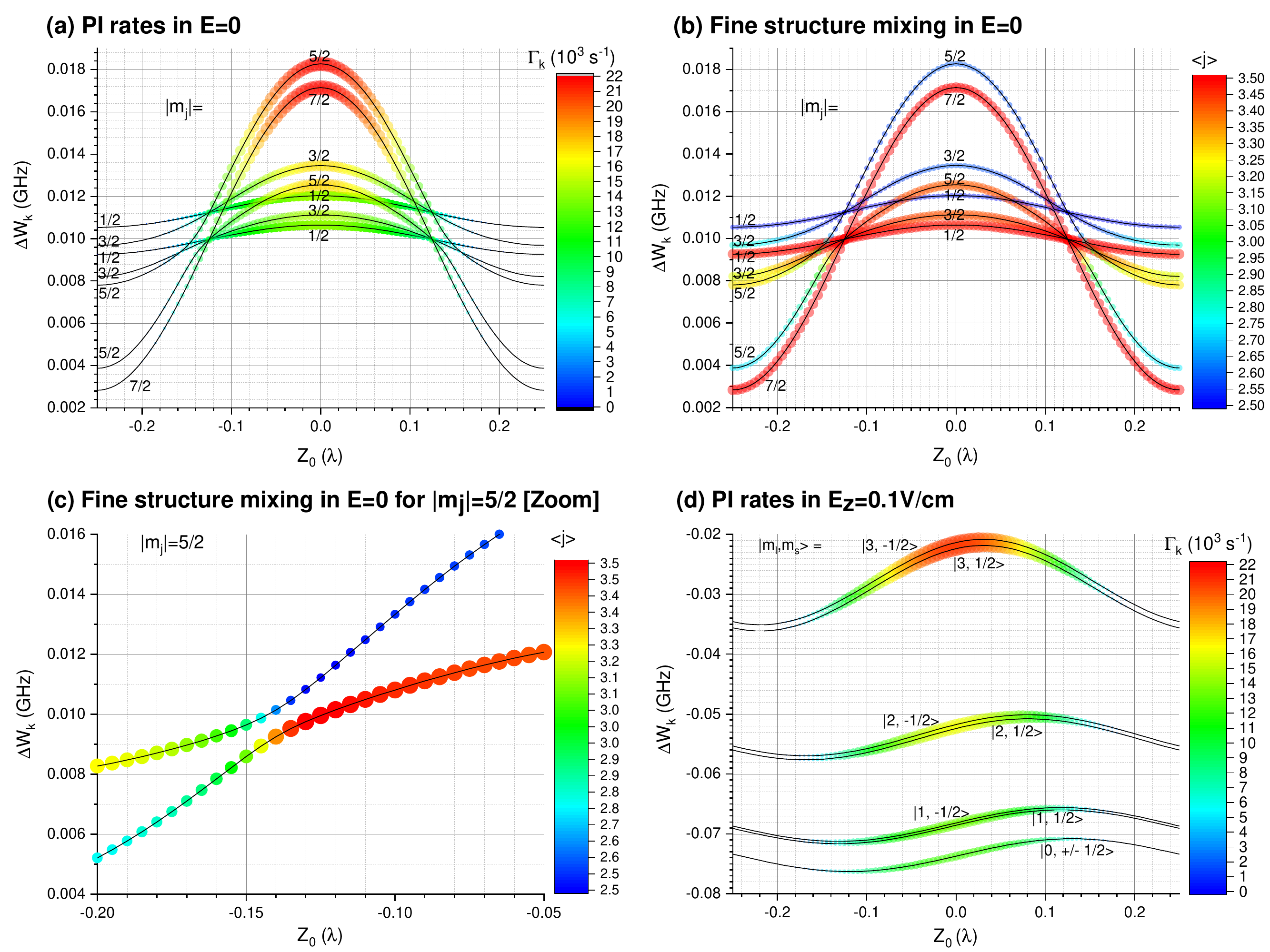}
\caption{(a) and (b): PECs of Rb $50F$ in an optical lattice with $\lambda=1064$~nm and a depth $2 V_0 = h \times 20$~MHz$ = 9867 E_{rec}$. The deviations of the PECs from the field-free Rb $50F_{7/2}$ state, $\Delta W$, in GHz are plotted vs CM position, $Z_{\rm 0}$, in units $\lambda$. The PEC labels show $|m_j|$. Symbol sizes and colors in (a) and (b) show PI rate and expectation value of $j$, respectively, on the given color scales. (c) Magnified view of a $|m_j|=5/2$ PEC-pair near the lattice inflection point, showing $j$-mixing and level repulsion.
(d) PECs for the same conditions as in (a-c), but with an added longitudinal DC electric field of 0.1~V/cm. The DC field breaks the fine-structure coupling, and the PECs correspond with position-independent adiabatic states $\vert 50F, m_\ell, m_s\rangle$. Symbol size and color show PI rate on the given color scale.
}\label{50D}
\end{centering}
\end{figure*}

In weak Rydberg-atom lattices it is $2 V_0 \ll 0.1 E_H/n^3$,
$\ell$-mixing plays no significant role \gfix{for states with $\ell<4$}, and PI is concentrated within a small number of non-mixed low-$\ell$ PECs that have large PI cross sections (see Fig.~1).
Hence, while the PI rate averaged over all PECs drops in proportion with lattice intensity, atoms on low-$\ell$ PECs may still photo-ionize at high rates.

Examples of PECs for $50F_j$ in a weak lattice with a depth of $2V_0 = h \times 20$~MHz$ = 9867 E_{rec}$ are shown in Fig.~\ref{50D}.
The $50F$-levels split into seven resolved components of conserved $m_j$. With the exception of $|m_j|=7/2$, there are two fine-structure states, $j=5/2$ and $7/2$, \hfix{that have a field-free splitting of 1.3~MHz and} that become mixed by the lattice. Solving Eqs.~\ref{H}-\ref{eq:gamma} in sub-spaces $\{ \vert 50F_{7/2}, m_j \rangle, \vert 50F_{5/2}, m_j \rangle \}$ yields the PECs and their PI rates. As seen in Fig.~\ref{50D}(a), the modulation depth of the PECs varies from strongly modulated at $|m_{j}|=7/2$ to barely modulated at $|m_{j}|=1/2$. The variation in PEC modulation depth arises from
the differing extent of the Rydberg-electron wavefunctions along the axis of the lattice, which results in varying amounts of averaging in Eq.~\ref{Vad}~\cite{Anderson.2012a}. Generally, the sublevels with lesser values of $|m_j|$ have wavefunctions that extend more in the direction of the lattice axis,
resulting in less deeply modulated PECs. Lattice-induced $j$-mixing is illustrated in Fig.~\ref{50D}(b), where the expectation value $j$ on some PECs varies considerably as a function of $Z_0$, while maintaining an average of 3 over pairs of coupled PECs with same $m_j$.
The fine structure coupling causes pairs of states of same $m_j$ to repel each other near the lattice inflection points at $Z_0=\pm \lambda/8$.  The level repulsion is seen best in Fig.~\ref{50D}(c), where we show a detailed view of the level pair with $m_j=5/2$ near a lattice inflection point.

As in Sec.~\ref{app2}, the PI rates $\Gamma_k(Z_0)$ generally
scale with the lattice intensity, which is $\propto(1+\cos( 2 k Z_0 ))$. Further, according to Eq.~\ref{sigmaxm}, the $\Gamma_k$-values at fixed $Z_0$ should increase with $m_\ell$, and by continuation, with $m_j$. This trend is obvious in Fig.~\ref{50D}(a). \hfix{The relative cross-section differences between taking coherent inner sums, as in Eq.~\ref{PIcross2}, and taking incoherent sums are 0.03, averaged over
all $\vert \psi_k(Z_0) \rangle$ in Fig.~3(a), with a standard deviation of 0.03.}

\hfix{To exhibit the $m_\ell$-dependence of cross sections and rates} more clearly, in Fig.~\ref{50D}(d) we show PECs and PI rates, $\Gamma_k$, with an additional longitudinal electric field along the $z$-direction.  The field is sufficiently strong to decouple the fine structure, but weak enough to not cause significant $\ell$-mixing with nearby $D$ and $G$ Rydberg states. The adiabatic states $\vert \psi_k \rangle$ associated with the PECs then approximately are $\vert 50F, m_\ell, m_s \rangle$. With orbital degeneracies lifted, the PECs follow from Eq.~\ref{Vad0} with $\psi({\bf{r}}_e) =
\langle {\bf{r}}_e \vert 50F, m_\ell \rangle$. There still is a small fine-structure splitting between PECs with same $m_\ell$ and different $m_s$, with the exception of $m_\ell = 0$, where the spin-up and -down states are \gfix{exactly degenerate}. The PI rates and their ratios between the PECs in Fig.~\ref{50D}(d), at a fixed $Z_0$, are now governed by Eq.~\ref{sigmaxm}, with $\ell=3$ and $\ell'=2$ or 4, \hfix{and the shell-averaged PI cross sections. The latter are $\bar{\sigma}_{50F}^{\epsilon', \ell'} = $650~barn for $\ell'=2$ and 3494~barn for $\ell'=4$.} Factoring in all dependencies in Eq.~\ref{sigmaxm} and (incoherently) summing the PI rates over $\ell'$,
the rates at the lattice intensity maxima, for conditions as in Fig.~\ref{50D}(d), vary between $21 \times 10^3~$s$^{-1}$ for $m_\ell = 3$
and $13 \times 10^3~$s$^{-1}$ for $m_\ell = 0$, \hfix{with no noticeable dependence on $m_s$.} In comparison, for the rates of black-body-radiation-induced bound-bound (Bbb) transitions and black-body photoionization (Bpi)~\cite{Traxler.2013,Anderson.2013b} we calculate
$\Gamma_{Bbb,50F} = 10.60 \times 10^3~$s$^{-1}$ and
$\Gamma_{Bpi,50F} =  0.77 \times 10^3~$s$^{-1}$, respectively, for a radiation temperature of 300~K and for all $m_\ell$. The lattice-induced PI should therefore be dominant over black-body-induced transitions.

% T= 300.0 K
% lqz  nqz  total_r(s^-1)  ion_r(s^-1)    up_r(s^-1)     down_r(s^-1) ion/total 
%  3  50    11370.686375    768.330399    3776.34965     6826.006319  

In potential experimental work, a lattice as in Fig.~\ref{50D} could be achieved, for instance, by focusing two counter-propagating $1064$-nm laser beams, with a power of 1~W each, into a confocal spot with $w_{\rm 0}=20~\mu$m. The PECs of Rb $nF$ states could then be studied via three-photon laser excitation from Rb $5S_{1/2}$. \gfix{A laser-spectroscopic measurement of PI-limited PEC widths of 50F states in lattices as in Fig.~\ref{50D} would require a laser linewidth  $\lesssim 1$~kHz.}

\section{Conclusion}
\label{Sec:conc}

%\rcfix{\st{We have obtained photoionization cross sections and decay rates of Rydberg atoms in plane-wave optical fields and in optical lattices. 
%We have noted that the electric-dipole approximation applies in
%Rydberg-atom photoionization, even though the atoms have a size on the order of the wavelength of the ionizing field. 
%We then proceeded to include photoionization into calculations of potential energy curves of Rydberg atoms in optical lattices. The degree of lattice-induced $\ell$-mixing has been related to the ratio between lattice depth and the intrinsic energy scale of the Rydberg atoms.  We have compared the significance of lattice-induced photoionization with that of natural decay and black-body effects.}}

\hfix{We have studied photoionization of Rb Rydberg atoms in an optical lattice formed by 1064-nm laser beams.} The strong Rydberg-atom lattices discussed in Sec.~\ref{app2} are suitable, for instance, for all-optical quantum initialization of high-angular-momentum states~\cite{Cardman.2020a} and other quantum-control applications. Weak Rydberg-atom lattices, as discussed in Sec.~\ref{app1}, are attractive for applications that include quantum computing and simulation~\cite{Zhang.2011a, Nguyen.2018,Barredo.2020}, and high-precision spectroscopy~\cite{Ramos.2017,Moore.2015a,Malinovsky.2020a}. Weak Rydberg-atom lattices at magic wavelengths~\cite{Safronova.2003a} can minimize trap-induced shifts of certain transitions~\cite{Moore.2015b, Ramos.2017}. Further, the $nF_j$ Rydberg states we have considered in our examples can serve as launch states for circular-state production~\cite{Ramos.2017,Cardman.2020a}. Some of these and other applications of Rydberg-atom optical lattices are subject to limitations from spectroscopic line broadening and decoherence caused lattice-induced photoionization. The photoionization rates as calculated in our paper will be useful in detailed feasibility estimates for these efforts.

%The relevance of the calculated potential energy curves and the photoionization rates of the associated adiabatic states in possible applications of Rydberg-atom optical lattices has been discussed, including in Rydberg-atom trapping, spectroscopy and quantum-state manipulation.

\begin{acknowledgments}

This work was supported by NSF Grant No. PHY-1806809 and NASA Grant No. NNH13ZTT002N.

\end{acknowledgments}

\appendix

\section{Atom-Field Interaction}
\label{atomfield}

In the Appendices we validate the electric-dipole approximation (EDA) in optical transitions  and photo-ionization (PI) of $\mu$m-sized Rydberg atoms with light. Expressions are extended to PI in an optical lattice. In the following, the ``$e$''-subscript on the relative electron coordinates, used in the main text, is dropped, and all lowercase coordinates are relative electron coordinates. 

The non-relativistic Hamiltonian for an $N$-electron atom with nuclear charge $Z$ is given by
\begin{equation}\label{hamiltonian}
\hat{H}=\sum_{i=1}^{N} \left( \frac{\hat{\textbf{p}}_{i}^{2}}{2m_e}-\frac{Z e^{2}}{4\pi\epsilon_{\rm 0}\hat{r}_{i}} \right) + \frac{1}{4\pi\epsilon_{\rm 0}} \sum_{i>j=1}^{N} \frac{e^{2}}{|\hat{\textbf{r}}_{i}-\hat{\textbf{r}}_{j}|}.
\end{equation}
\noindent The first sum includes the kinetic and potential energy of each electron in the Coulomb field of the nucleus, and the second the electrostatic repulsion between pairs of electrons. The interaction of the atom with an electromagnetic field can be taken into account by replacing $\hat{\textbf{p}}_{i}$ with $\hat{\textbf{p}}_{i}+ e \textbf{A}(\hat{\textbf{r}}_{i},t)$, where $\textbf{A}(\hat{\textbf{r}}_{i},t)$ is the vector potential. The resulting interaction added to Eq.~\ref{hamiltonian} is
\begin{multline*}
\hat{H}_{\rm 1}=\sum_{i=1}^{N} \lbrace\frac{e}{2m_e}[\hat{\textbf{p}}_{i} \cdot \textbf{A}(\hat{\textbf{r}}_{i})+\textbf{A}(\hat{\textbf{r}}_{i}) \cdot \hat{\textbf{p}}_{i}] \\ +\frac{e^{2}}{2m_e}\textbf{A}^{2}(\hat{\textbf{r}}_{i})\rbrace.
\end{multline*}
The $\textbf{A}^{2}(\hat{\textbf{r}}_{i})$ term gives rise to the ponderomotive potential that is responsible for the trapping of Rydberg atoms in an optical lattice~\cite{Dutta.2000a, Zhang.2011a}.
In a QED treatment, the Feynman diagram of the $\textbf{A}^{2}(\hat{\textbf{r}}_{i})$ term is a vertex with two instantaneous photons~\cite{Sakuraibook}.
%In low-intensity electromagnetic radiation fields, the $\textbf{A}^{2}(\hat{\textbf{r}}_{i})$ term can be neglected.
The $\hat{\bf{A}} \cdot \hat{\bf{p}}$-term causes a wide range of atom-field interactions, including light-induced and black-body-radiation-induced PI.
In the Coulomb gauge, $\nabla \cdot \textbf{A}=0$, the operators $\hat{\textbf{p}}_{i}$ and $\textbf{A}(\hat{\textbf{r}}_{i})$ commute, and the $\hat{\bf{A}} \cdot \hat{\bf{p}}$ interaction writes
\begin{equation*}
\hat{H}_{\rm int}=\sum_{i}\biggl(\frac{e}{m_e}\textbf{A}(\hat{\textbf{r}}_{i}) \cdot \hat{\textbf{p}}_{i}\biggr).
\end{equation*}
\noindent In the present work we consider a Rydberg atom with one active electron. In this case, the sum can be dropped, and the position and momentum operators $\hat{\textbf{r}}$ and $\hat{\textbf{p}}$ are just for the Rydberg electron. In a source-free field, the electric field and vector potential are related by $\textbf{E}=-(\partial \textbf{A}/\partial t$)~\cite{Jacksonbook}. We consider a linearly polarized plane wave with electric-field amplitude $E_0$, and choose the $x$-axis in propagation and the $z$-axis in field direction,  $\textbf{A}(\textbf{r},t)=\frac{E_0}{2i\omega}\hat{\bf{z}}e^{i(k x-\omega t)} + cc$. There, $\omega$ is the angular frequency and $k$ the wavenumber.

The matrix element $\langle f | \hat{H}_{\rm int} | i \rangle$ is, in the rotating frame~\cite{Friedrichbook, Bethebook},
\begin{equation}\label{Hint}
\langle f | \hat{H}_{\rm int} | i \rangle=-\frac{ e \hbar E_0}{2m_e \omega}\int\psi^{\ast}_{f} e^{ikx}\frac{\partial}{\partial z}\psi_{i}\medspace d^{3}r,
\end{equation}
where $|i\rangle$ and $|f\rangle$ are the initial and final states with wavefunctions $\psi_{i}$ and $\psi_{f}$. Using Fermi's golden rule, the transition rate is
\begin{equation}\label{Rate}
\Gamma=\frac{2\pi}{\hbar}|\langle f|\hat{H}_{\rm int}|i\rangle|^{2}\rho(\epsilon) \quad,
\end{equation}
with the density of states $\rho(\epsilon)$ at the final-state energy. The rates are proportional to the intensity, regardless of whether the EDA, which amounts to setting $e^{ikx} = 1$, can be made or not.

To compute the matrix elements $M_{A}$ = $\int\psi^{\ast}_{f} e^{ikx}\frac{\partial}{\partial z}\psi_{i}\medspace d^{3}r$ \hfix{for unperturbed, fine-structure-free Rydberg states} we use the usual notations $\psi_{n,\ell,m_\ell}(r,\theta,\phi)$ = $R_{n,\ell}(r)Y_{\ell}^{m_\ell}(\theta,\phi)$~\cite{Bethebook}, and $R_{n,\ell}(r)$ = $u_{n,\ell}(r)/r$. The quantum numbers $(n,\ell,m_\ell)$ and $(n',\ell',m'_\ell)$ are for the initial and final states, respectively. The radial wavefunctions are calculated according to Ref.~\cite{Reinhard.2007a}, using model potentials from Ref.~\cite{Marinescu.1994}. The Jacobi-Anger relation~\cite{Arfkenbook},
\begin{equation*}
e^{i a \cos\phi}=\sum_{\widetilde{m}=-\infty}^{\infty}i^{\widetilde{m}}J_{\widetilde{m}}(a)e^{i\widetilde{m}\phi},
\end{equation*}
\noindent expresses $e^{ikx}$ as an azimuthal Fourier series. The matrix element $M_A$, including both angular and radial factors, then becomes
\begin{widetext}
\begin{multline}\label{matel}
M_A=i^{m'_\ell-m_\ell}\frac{1}{2}\sqrt{\frac{2\ell'+1}{2\ell+1}\frac{(\ell'-m_\ell')!}{(\ell'+m_\ell')!}\frac{(\ell-m_\ell)!}{(\ell+m_\ell)!}} \\ \times
\Biggl\lbrace \int u_{n',\ell'}(r)[u'_{n,\ell}(r)-\frac{u_{n,\ell}(r)}{r}(\ell+1)] \left[ \int J_{m_\ell'-m_\ell}(k r \sin\theta) P_{\ell'}^{m_\ell'}(\cos\theta)P_{\ell+1}^{m_\ell}(\cos\theta)(\ell-m_\ell+1)\sin\theta \medspace d\theta \right] dr \\
+ \int u_{n',\ell'}(r)[u'_{n,\ell}(r)+\frac{u_{n,\ell}(r)}{r}\ell]\left[\int J_{m_\ell'-m_\ell}(k r \sin\theta) P_{\ell'}^{m_\ell'}(\cos\theta)P_{\ell-1}^{m_\ell}(\cos\theta)(\ell+m_\ell)\medspace \sin\theta \medspace d\theta \right] dr \Biggr\rbrace.
\end{multline}
\end{widetext}
\noindent For PI the transitions are from a bound to a free state. In this case, the radial wavefunction $u_{n',\ell'}$ is replaced by a free radial wavefunction $u_{\epsilon', \ell'}$. The free radial wavefunctions are normalized in energy, $\int u_{\epsilon', \ell'}(r)u_{\epsilon, \ell'}(r)\medspace dr=\delta(\epsilon-\epsilon')$, and the density of states $\rho(\epsilon) = 1$.

\section{General behavior of the matrix elements}\label{noapprox}

\begin{figure*}[t]
\begin{centering}
\includegraphics[scale=0.36]{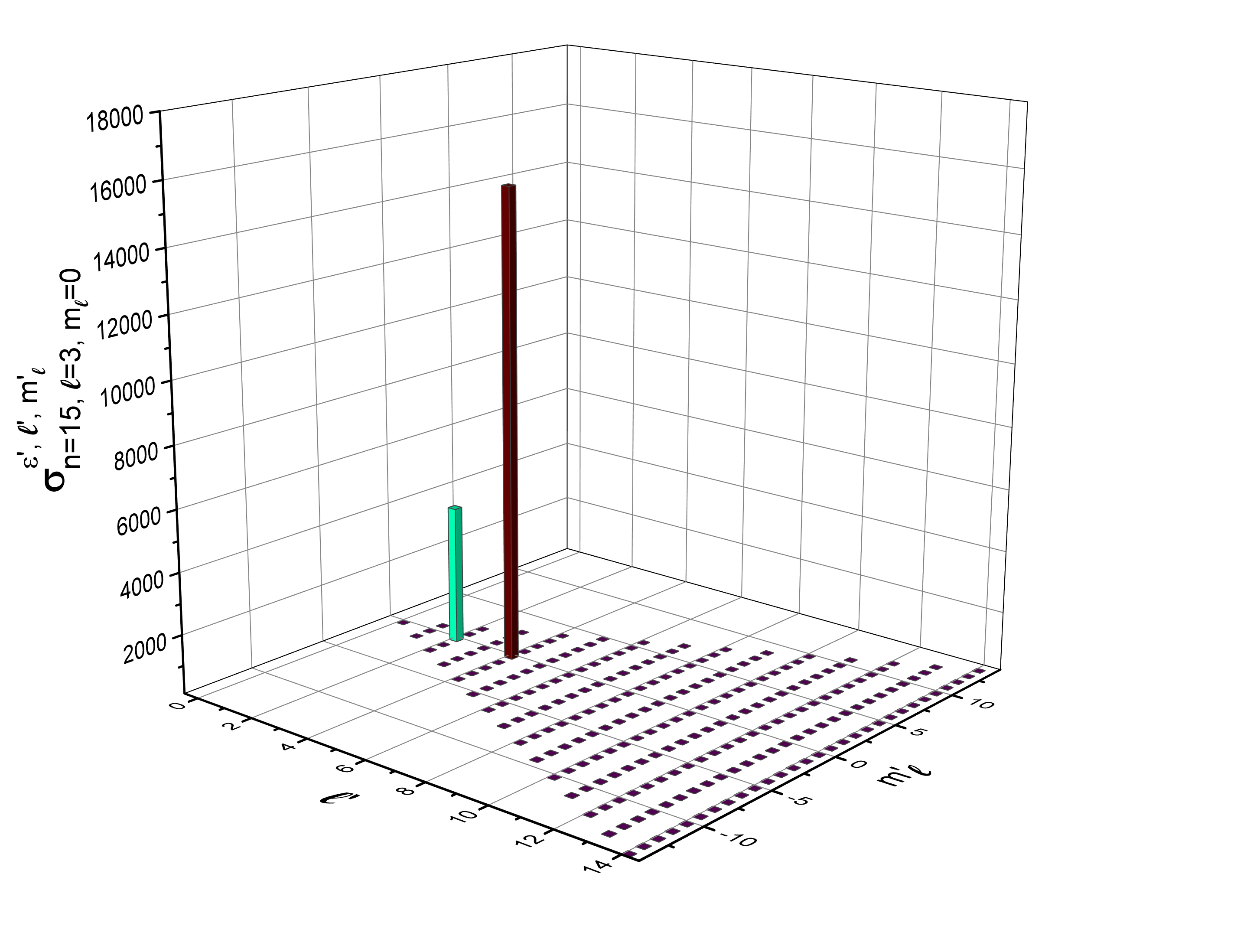}
\includegraphics[scale=0.36]{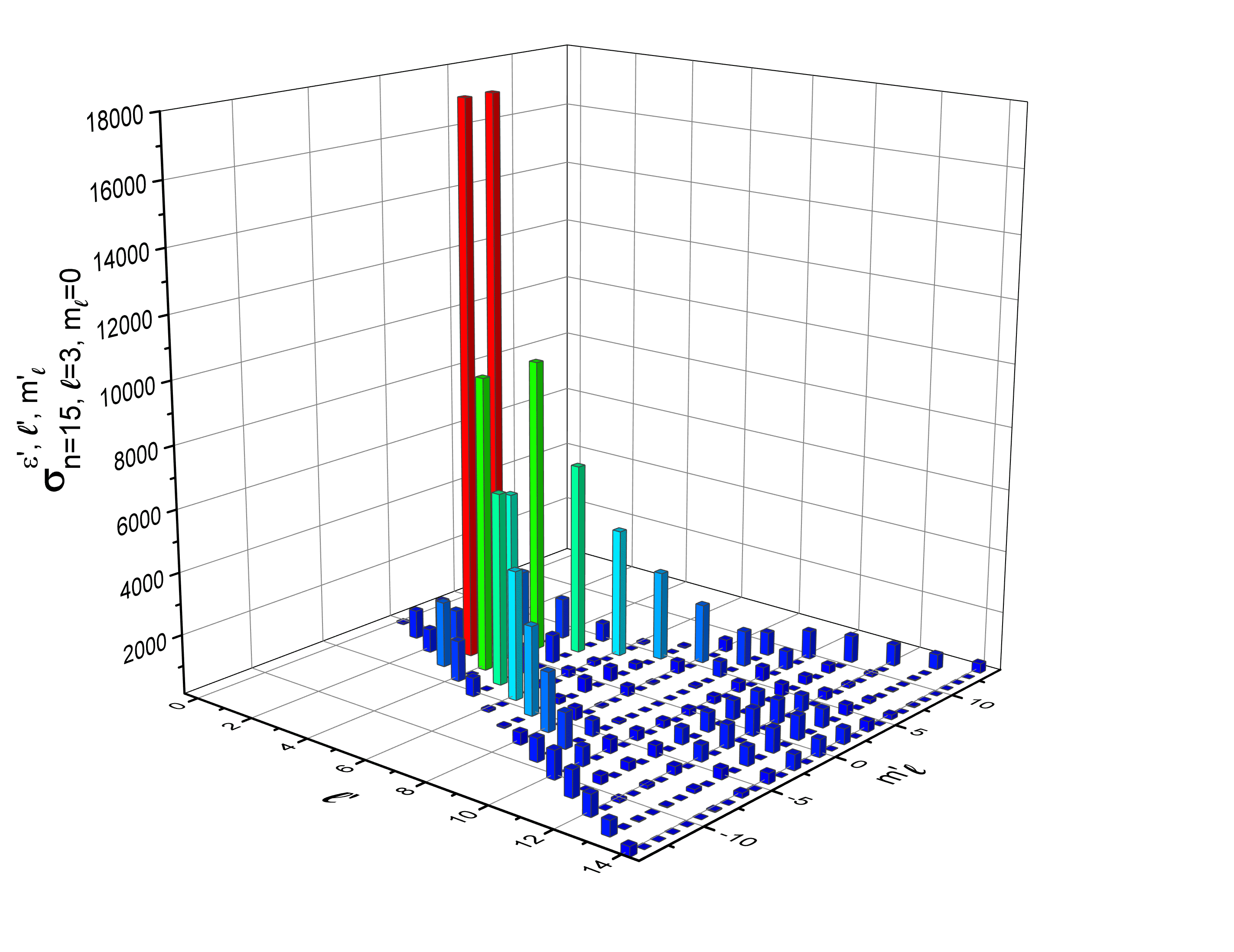}
\caption{(a) Cross sections for PI of the Rb $|n=15,\ell=3,m=0\rangle$ state
with $z$-polarized 532-nm light for transitions into the continuum states $|\epsilon'=0.08342, \ell', m'_\ell \rangle$, plotted for a range of values of final $\ell'$ and $m'_\ell$. In the calculations the EDA is not applied.
The only transitions with matrix elements of non-negligible amplitude are the electric-dipole-allowed transitions. (b)
PI cross sections for the same transitions as in (a), but with the wavelength of the field artificially reduced by a factor of $\kappa=1000$. Transitions that violate the electric-dipole selection rules now have larger values, often exceeding those of the two electric-dipole-allowed transitions.}\label{fig}
\end{centering}
\end{figure*}

The range of relevant PI channels, {\sl{i.e.}} the range of the $\ell'$ and $m'_\ell$ quantum numbers for which the matrix elements $M_A$ for a given initial state are large, largely depends on the magnitude of the Bessel-function arguments. The EDA, $e^{ikx}=1$, corresponds with $J_{m_\ell'-m_\ell}(k r \sin\theta) = \delta_{m_\ell',m_\ell}$. Here we assess how well the EDA applies to Rydberg-atom PI with light. At first glance, one may suspect the EDA to be invalid because $kr \sim 1$.

\subsection{Before making the EDA}\label{without}

Equation~\ref{matel} yields a selection rule that arises from the three functions within the $\theta$ integrals (one Bessel function and two associated Legendre functions), which all have well-defined parity about $\pi/2$. Considering the parity behavior of the associated Legendre functions with $\ell$ and $m_\ell$, and noting that the Bessel function terms are always even, we find the selection rule that $\ell+m_\ell+\ell'+m'_\ell+1$ must be even (meaning that about half of the transitions out of a state with given $\ell$ and $m_\ell$ are allowed).

In the limit $ k r \rightarrow 0$ (as when the  EDA is valid),
$J_{m_\ell'-m_\ell}(k r \sin\theta) = \delta_{m_\ell',m_\ell}$. The orthogonality of the Legendre functions
then yields the usual  (very restrictive) electric-dipole selection rules $m'_\ell - m_\ell=\Delta m_\ell =0$ (for $z$-polarized light) and $\ell' - \ell = \Delta\ell=\pm1$.

The cross section $\sigma$,  the rate $\Gamma$, and the light intensity $I$ follow
\[
\sigma = \hbar \omega \Gamma / I \quad ,
\]
which after insertion of Eqs.~\ref{Hint} and~\ref{Rate} yields
\begin{equation}\label{PIMdz}
\sigma_{z, n, \ell, m_\ell}^{\epsilon', \ell', m'_\ell}=\frac{\pi e^{2}\hbar^{2}}{\epsilon_{\rm 0} m_{\rm e}^{2} \omega c} \left|M_A \right|^{2}\left(\frac{1}{E_H \it a_{\rm 0}^{\rm 2}}\right) \quad.
\end{equation}
The result is in SI units, m$^2$, the matrix element $M_A$ in atomic units, according to  Eq.~\ref{matel}, and the term in () converts $ |M_A |^2$  from atomic into SI units. To illustrate the typical PI behavior of Rydberg atoms in light fields, we calculate matrix elements and cross sections
following Eqs.~\ref{matel} and~\ref{PIMdz} for PI of a Rb Rydberg atom by 532-nm light.
In Fig.~\ref{fig}(a), we display $\sigma_{z, n, \ell, m_\ell}^{\epsilon', \ell', m'_\ell}$ for PI of Rb $|n=15,\ell=3,m=0 \rangle$ to the continuum states $|\epsilon'=0.08342, \ell', m'_\ell \rangle$. \rcfix{We choose the $n=15$ Rydberg state instead of $n=50$ because it requires less computing power while still allowing us to explain the validity of the EDA.} It is seen that the only transitions that have a non-negligible PI cross section are the electric-dipole-allowed transitions in the assumed $z$-polarized light, $\Delta m_\ell =0$ and $\Delta\ell=\pm 1$. The weaker of the two electric-dipole-allowed PI channels is into $|\epsilon'=0.08342, \ell'=2, m'_\ell= 0 \rangle$ and has a cross section of 4483~barn. The strongest electric-dipole-violating channel is into  $|\epsilon'=0.08342, \ell'=5, m'_\ell=\pm 1 \rangle$ and has a calculated cross section of 0.18~barn, which is smaller than that of the weaker electric-dipole-allowed channel by a factor of $4 \times 10^{-5}$. It thus appears the EDA applies exquisitely well to Rydberg-atom PI by light.

% 0.18096/4483 = 4.036 e-5
% l' m' sigma
% 4	0	15218.50684
% 2	0	4483.61897
% 5	1	0.18096 0
% 5	-1	0.18096

% fin=500, rminp=1d-5m
% n, l, m = 15, 3, 0
% eps', l', m' = 0.083418, 4, 0
% outfile = test1.dat
% lfree mfree      rmati1d        sigi1d(barn)
%  4     0    -0.1484124309736     15218.507

\begin{figure*} [t]
\begin{centering}
\includegraphics[scale=0.35]{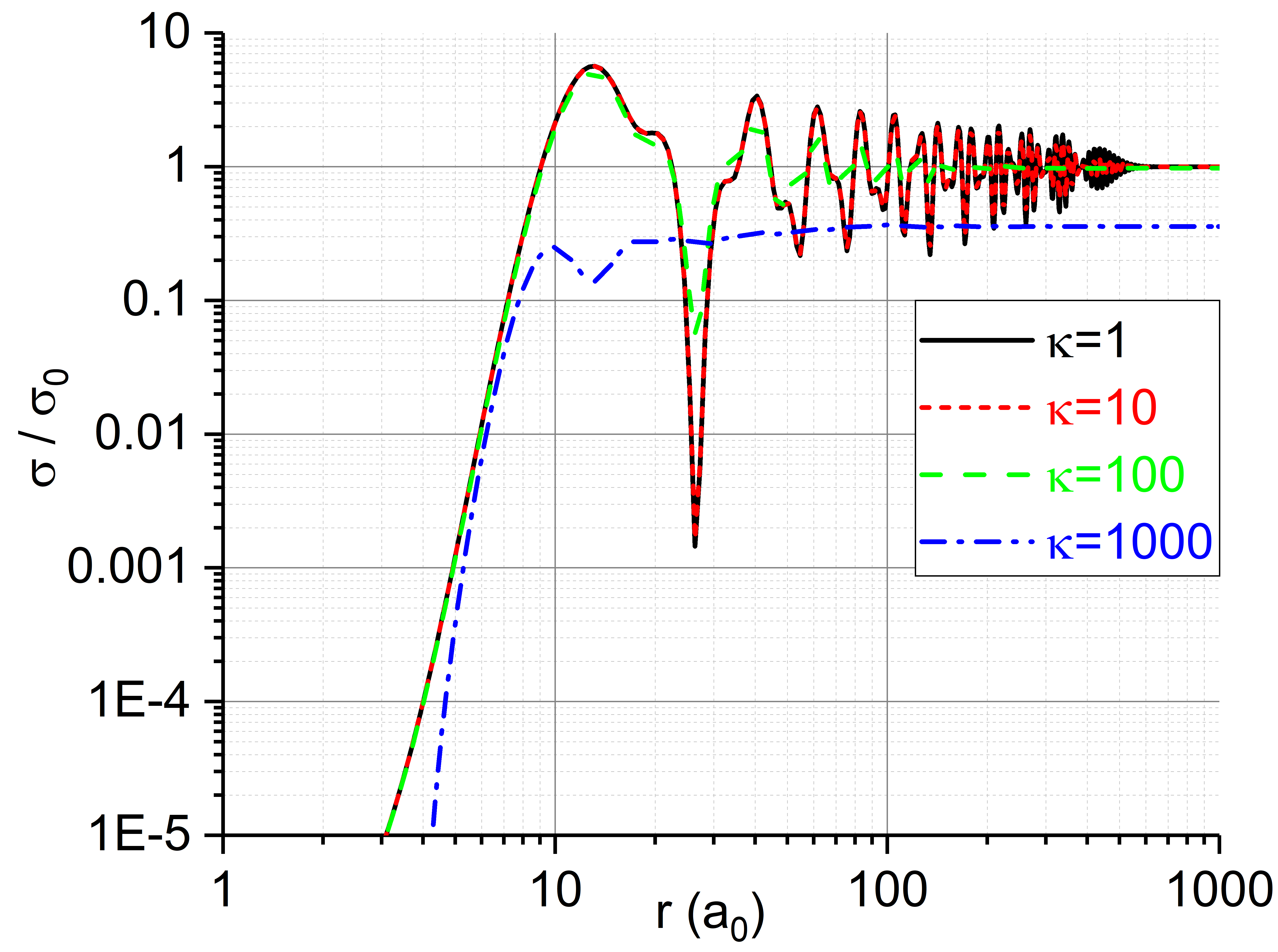}
\includegraphics[scale=0.35]{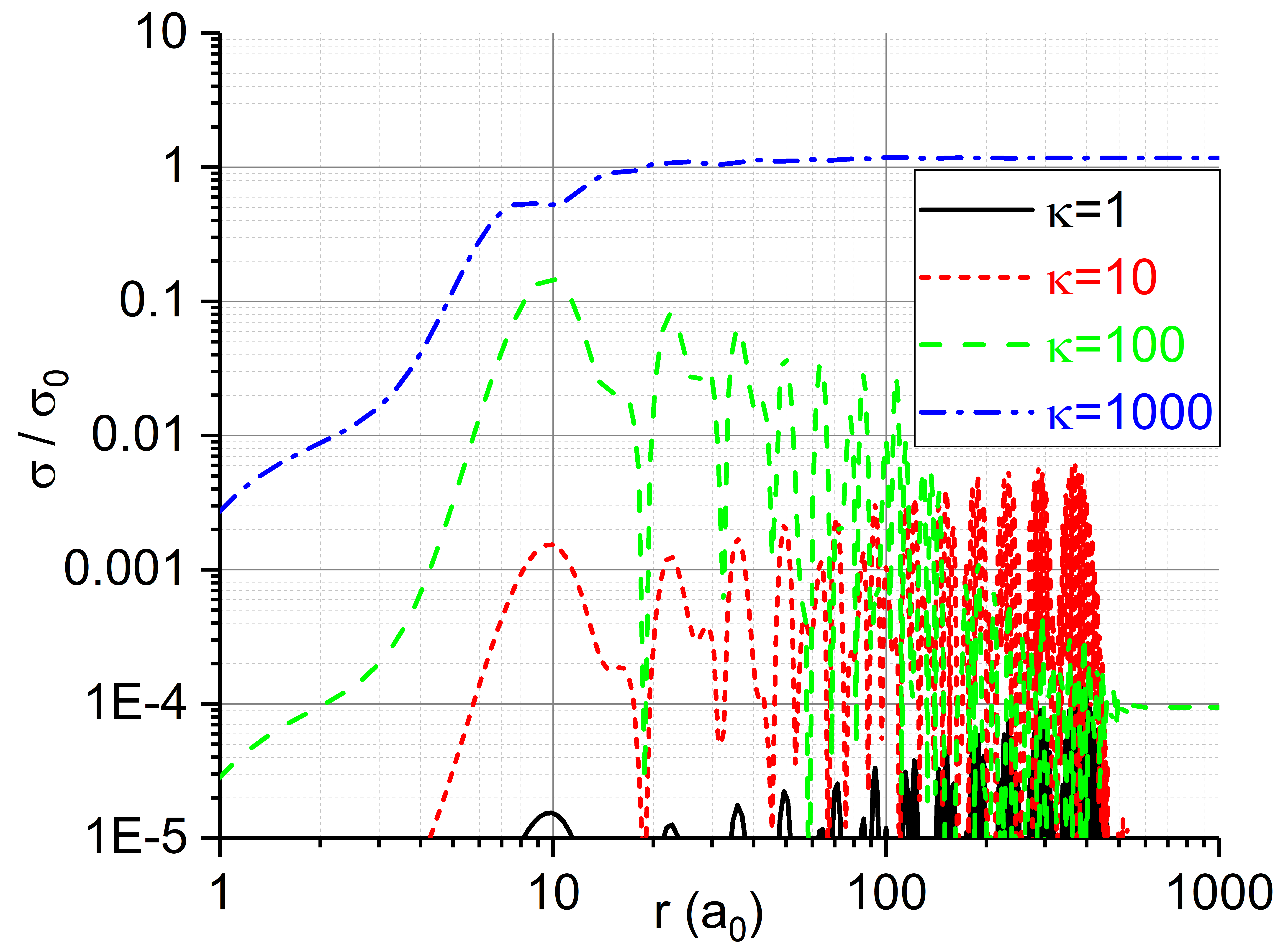}
\caption{Cross sections in units $\sigma_0 = 15220$~barn
for PI of the Rb $|n=15,\ell=3,m=0\rangle$ state
with $z$-polarized 532-nm light for the electric-dipole-allowed transition into the continuum state $|\epsilon'=0.08342, \ell'=4, m'_\ell=0\rangle$ (left), and for the electric-dipole-violating transition into the continuum state $|\epsilon'=0.08342, \ell'=3, m'_\ell=1\rangle$ (right), as a function of the upper integration limit in the \gfix{matrix-element} calculation, $r$, and for the indicated parameters $\kappa$ (see text).}\label{M}
\end{centering}
\end{figure*}

The strong validity of the EDA for Rydberg-atom PI may appear somewhat unexpected, because
both initial and final states have sizes on the order of or exceeding the optical wavelength, and the usual argument made when invoking the EDA, namely that $e^{ikx}=1$ within
the atomic volume, is actually not valid.
To expose conditions under which electric-dipole-violating transitions would be important, we increase the wavenumber $k$ in the $e^{ikx}$ phase factor (and in the Bessel function argument in Eq.~\ref{matel}) by an artificial factor $\kappa$, so as to artificially enhance EDA-violation, while holding everything else fixed (including the energy of the continuum state). While this
is physically not possible, the numerical exercise allows us to explore where the unexpected validity of the EDA arises from when performing the integration in Eq.~\ref{matel}. By increasing the argument of the Bessel functions by $\kappa$, we artificially increase the variation of the Bessel functions in the matrix-element integration. Cross sections for PI of $|n=15,\ell=3,m=0\rangle$ to the continuum states $|\epsilon'=0.08342, \ell', m'_\ell \rangle$ calculated with $\kappa=1000$ are shown in Fig.~\ref{fig}(b). The EDA is evidently not valid any more, as a large number of electric-dipole-forbidden transitions occur.  The strongest electric-dipole-allowed channel now is to $|\epsilon'=0.08342, \ell'=4, m_\ell =0\rangle$, with a cross section of 5451~barn, while the strongest EDA-violating channel is to
$|\epsilon'=0.08342, \ell'=3, m_\ell = \pm 1 \rangle$, with a cross section of 17850~barn.
Fig.~\ref{fig}(b) also shows a ``checker board'' pattern, which reflects the selection rule that $\ell+m_\ell+\ell'+m'_\ell+1$ must be even (which still holds for EDA-violating transitions).

%Maybe drop this
%\rcfix{The atomic size of the $n=50$ state  is only about a factor of 10 larger than that for $n=15$ used in this Appendix. Such an increase in size relative to the laser wavelength allows the EDA to hold for the $n=50$ level under study in the main text, for the arguments of the Bessel functions are still sufficiently small.}

% l' m' sigma
% 3	1	17851.8332
% 3	-1	17851.8332
% 4	-2	9479.41363
% 4	2	9479.41363
% 5	-3	6182.31364
% 5	3	6182.31364
% 4	0	5451.64477

For more insight, in Fig.~\ref{M} we plot the cross sections for a few cases
of PI of $\vert n=15, \ell=3, m_\ell=0 \rangle $ with 532-nm light and the indicated values of $\kappa$ as a function of cut-off radius of the radial integration in Eq.~\ref{matel}.
Considering the physical case first, for which $\kappa=1$, we find that the \gfix{matrix element} of the EDA-allowed transition integrates close to its final value already within a radius of about 50~$a_0$ and then oscillates around the final value, with the oscillations damping away in the outer reaches of the atomic volume.
The oscillations originate from the structure of bound- and free-state wavefunctions.
The effective range of the atom-field interaction appears to be confined to $r \lesssim 50~a_0$. One may say that the Rydberg atom tends to photoionize close to its center, a finding that is in accordance with calculations performed elsewhere~\cite{Giusti-Suzor.1987}. Since \gfix{the matrix element} integrates close to \gfix{its final value} within a volume that is indeed much smaller than the physical wavelength, the phase variation of the field in the outer regions of the atom, $r \gtrsim 50~a_{\rm 0}$, becomes irrelevant, making the EDA applicable even though the atom diameter is on the order of the optical wavelength.

The results in Fig.~\ref{M} for artificially reduced wavelength, {\sl{i.e.}} with the wavenumber $k$ in the $e^{ikx}$ phase factor multiplied with a $\kappa >1$, show that substantial changes of the cross sections from their physical values require $\kappa$-values approaching 1000, corresponding to effective wavelengths in the phase factor (and the Bessel-function arguments in Eq.~\ref{matel}) as low as several tens on $a_0$. In that case, the phase of the field does vary substantially over the volume within which the physical, $\kappa$=1-matrix \gfix{element integrates to near its asymptotic value}. For $\kappa$ approaching 1000, the EDA breaks down, leading to substantial changes of the cross sections of electric-dipole-allowed PI channels, as well as to the emergence of large cross sections in dipole-forbidden PI channels.
We conclude that the validity of the EDA is linked to the behavior that the physical matrix \gfix{elements integrate} to near their asymptotic values within a small volume of only several tens of $a_0$ in radius around the atomic center. The oscillations in the integrals in Fig.~\ref{M} that occur outside that volume are inconsequential, as they damp out. Hence, it is sufficient for the field phase in $e^{ikx}$ to be flat over a volume of just several tens of $a_0$ in radius, regardless \gfix{of} how large the atom is, for the EDA to be valid. This very relaxed condition reflects the somewhat surprising validity of the EDA for PI of Rydberg atoms with light.

\subsection{With the electric dipole approximation}

Making the EDA by setting $kr=0$ in Eq.~\ref{matel}, one finds for the matrix elements relevant to the main text of this paper
\begin{widetext}
\begin{equation}\label{matel_approx}
M_A = \sqrt{\frac{(\ell_{>}+m_\ell)(\ell_{>}-m_\ell)}{(2\ell_{>}+1)(2\ell_{>}-1)}} \times \left\{
  \begin{array}{l l}
    \int u_{n',\ell'}(r)[u'_{n,\ell}(r)-\frac{u_{n,\ell}(r)}{r}\ell_{>}] \medspace dr & \quad \text{if $\ell_{>}=\ell'=\ell+1$}\\
    \int u_{n',\ell'}(r)[u'_{n,\ell}(r)+\frac{u_{n,\ell}(r)}{r}\ell_{>}] \medspace dr & \quad \text{if $\ell_{>}=\ell=\ell'+1$}
  \end{array} \right. ,
\end{equation}
\end{widetext}
with the usual electric-dipole selection rules for the changes in angular-momentum quantum numbers for $z$-polarized light, $\Delta \ell = \pm 1$ and $\Delta m_\ell =0 $. The full interaction matrix element then is
\begin{equation*}\label{velform}
\langle f | \hat{H}_{\rm int} | i \rangle=-\frac{|e| \hbar E_0}{2 m_e \omega} M_A \quad ,
\end{equation*}
with expressions for the resultant PI rates still given by Eq.~\ref{Rate}. For linearly polarized light with arbitrary polarization direction $\hat{\bf{n}}$,
\begin{equation*}\label{velform1}
M_A = \hat{\bf{n}} \cdot \int\psi^{\ast}_{f} \medspace \frac{\hbar}{i} \nabla\medspace\psi_{i}\medspace d^{3}r \quad ,
\end{equation*}

These forms of the matrix elements
are known as the ``velocity'' form. With the EDA valid, the matrix elements can be expressed in other forms using commutation relations between operators. The relation $[\hat{\textbf{r}}, \hat{H}_{\rm 0}] = \frac{i\hbar}{m}\hat{\textbf{p}}$, which applies to systems with field-free Hamiltonians $\hat{H}_{\rm 0}$ that have momentum-independent potentials~\cite{Bethebook}, allows the matrix elements to be written in terms of the position operator. In this form, known as  ``length'' form, it is
\begin{equation}\label{Hint_approx}
\langle f| \hat{H}_{\rm int}|i\rangle=\frac{e E_0}{2}\hat{\textbf{n}}\cdot\int \psi_{f}^{\ast} \medspace \textbf{r} \medspace \psi_{i} \medspace d^{3}r \quad ,
\end{equation}
\noindent with the commonly used dipole matrix element $M_{\rm A, r}=\hat{\textbf{n}}\cdot\int \psi_{f}^{\ast} \medspace \textbf{r} \medspace \psi_{i} \medspace d^{3}r$~(see, for instance, \cite{Bethebook} equation 60.7f).

Finally, if the potential in $\hat{H}_{\rm 0}$ is a Coulomb potential, the matrix elements may be expressed in ``acceleration'' form, in which the commutation relation $[\hat{\textbf{p}}, \hat{H}_{\rm 0}]=-i\hbar \nabla \hat{V_0}$
with atomic potential $\hat{V_0}$
is used to express the matrix elements in terms of the Coulomb acceleration $(Z \hat{\textbf{r}})/r^{3}$~\cite{Bethebook}. In the length, velocity, and acceleration forms, the matrix elements accumulate to their asymptotic values at large, intermediate, and small values of $r$, respectively~\cite{Bethebook, Giusti-Suzor.1987, Muller.1993}. In the
present work, the velocity form, the most-generally valid form, must be
used because it allows for $\ell$-dependent
model potentials with non-Coulombic corrections~\cite{Marinescu.1994}, which is what we use in the computation of the wavefunctions. We have checked that length- and velocity forms yield identical results for high $\ell$, where the model potential becomes $\ell$-independent. Even at $\ell=0$, the worst case, the length- and velocity forms yield PI cross sections that differ by less than 15$\%$. For completeness it is further noted that for bound-bound microwave transitions of Rydberg atoms the length form is generally acceptable, because in that case the matrix elements are dominated by the outer reaches of the Rydberg wavefunctions, where the model potentials are essentially $\ell$-independent.

\section{PI in an optical lattice}\label{olat}

In a one-dimensional optical lattice formed by two counter-propagating beams with equal field amplitude $E_0$, polarization along the $z$-axis, and with beams propagating (anti)parallel with the $x$-axis, the electric field  is
\[ {\bf{E}} = \hat{\bf{z}} E_0 \Big[ \cos\big\{ k (x-X_0) - \omega t \big\} +
\cos\big\{ - k (x-X_0) - \omega t \big\} \Big]  \quad,
\]
where $X_0$ denotes the center-of-mass (CM) displacement of the atom from an intensity anti-node of the lattice. It is then found that the matrix element to be used in place of Eq.~\ref{matel}, with the EDA not being made, is
\begin{widetext}
\begin{eqnarray}\label{matelolat}
M_A & = &i^{m'_\ell-m_\ell}\frac{1}{2}\sqrt{\frac{2\ell'+1}{2\ell+1}\frac{(\ell'-m_\ell')!}{(\ell'+m_\ell')!}\frac{(\ell-m_\ell)!}{(\ell+m_\ell)!}} \nonumber \\
~ & \times &
\Biggl\lbrace \int u_{n',\ell'}(r)[u'_{n,\ell}(r)-\frac{u_{n,\ell}(r)}{r}(\ell+1)] \left[ \int J_{m_\ell'-m_\ell}(k r \sin\theta) P_{\ell'}^{m_\ell'}(\cos\theta)P_{\ell+1}^{m_\ell}(\cos\theta)(\ell-m_\ell+1)\sin\theta \medspace d\theta \right] dr
\nonumber \\
~& ~ & + \int u_{n',\ell'}(r)[u'_{n,\ell}(r)+\frac{u_{n,\ell}(r)}{r}\ell]\left[\int J_{m_\ell'-m_\ell}(k r \sin\theta) P_{\ell'}^{m_\ell'}(\cos\theta)P_{\ell-1}^{m_\ell}(\cos\theta)(\ell+m_\ell)\medspace \sin\theta \medspace d\theta \right] dr \Biggr\rbrace \nonumber \\
~ & \times &
\Biggr\{
\begin{array}{cc}
2 \cos(k X_0), & m'_\ell - m_\ell \quad {\rm even} \\
~ & ~\\
2 i \sin(k X_0), & m'_\ell - m_\ell \quad {\rm odd}
\end{array} \quad .
\end{eqnarray}
\end{widetext}
The PI rates scale with $| M_A |^2$. For even $m'_\ell - m_\ell$, the rates are proportional to the lattice-field intensity, which
is $4 I [\cos(k X_0)]^2$, with $I$ denoting the intensity of a single lattice beam, while for odd $m'_\ell - m_\ell$ the  rates scale with the derivative-square of the lattice electric field along the $x$-direction.

In the analysis performed in this Appendix we have assumed a light polarization pointing along $z$ and 
a field propagating along $x$, because this allows for a transparent evaluation of the matrix elements in
the general case that the EDA does not apply.
Now we have established that the EDA applies, for the physics presented here. It follows that only the electric-dipole-allowed case $m'_\ell - m_\ell = 0$ in Eq.~\ref{matelolat} is relevant. The equation greatly simplifies and takes the form
of Eq.~\ref{matel_approx}, with an $X_0$-dependent term $2 \cos(k X_0)$ multiplied on it. In essence this means that, if the EDA applies, as in our case, the PI cross section in an optical lattice is the same as in a plane wave, and that the field intensity to be used for computing the PI rate from this cross section is the field intensity at the CM location of the atom. \gfix{If the EDA were substantially violated (which is not the case),} electric-dipole-forbidden transitions with odd $m'_{\ell}-m_{\ell}$ \gfix{would, in principle, become} allowed, and the PI rates following from $M_{A}$ \gfix{would  not generally be proportional to intensity at the atomic CM location.} In that case, the usual concept of a PI cross section \gfix{would become}, fundamentally, invalid. 

In the main text of this paper, it is more convenient to assume an atomic quantization axis along $z$, one-dimensional optical-lattice laser beams propagating along $z$, and laser polarization along $x$. This allows us to take advantage of azimuthal symmetry in the calculation of the PECs of the lattice, substantially reducing the computational effort. 
The Rydberg-atom CM position, denoted $X_0$ in the Appendix, turns into $Z_0$ in the main text of the manuscript.  

%Further, the matrix elements $M$ in the main text are radial matrix elements, with the angular ($m_\ell$-dependent) parts factored out.

%\begin{singlespace} % Bibliography must be single spaced
\bibliographystyle{apsrev}
\bibliography{References}   % Use the BibTeX file ``References.bib''.
%\end{singlespace}

\end{document}